\documentclass[reprint,amsmath,amssymb,aps]{revtex4-2}

\usepackage[T1]{fontenc}
\usepackage{graphicx}
\usepackage{subfigure}
\usepackage{multirow}
\usepackage{xcolor,soul}
\usepackage{longtable}
\usepackage{dcolumn}
\usepackage[colorlinks=true,allcolors=blue]{hyperref}
\usepackage{bm}
\usepackage{amsfonts}
\usepackage{amsmath}
\usepackage{amssymb}
\usepackage{svg}
\usepackage{amsthm}
\usepackage{MnSymbol}
\usepackage{enumerate}
\usepackage{enumitem}

\theoremstyle{definition}

\begin{document}

\title{Minimal ancilla for boosting type-I photonic fusion}

\author{A.\,A.\,Melkozerov$^{1,2}$}
\email{melkozerov.alex@gmail.com}
\author{S.\,S.\,Straupe$^{3,1,2}$}
\author{M.\,Yu.\,Saygin$^{3,2}$}

\affiliation{\it $^1$ Russian Quantum Center, Bolshoy Boulevard 30, building 1, Moscow, 121205, Russia \\
\it $^2$ Faculty of Physics, M.\,V. Lomonosov Moscow State University, Leninskie Gory 1, Moscow, 119991, Russia \\
\it $^3$ Sber Quantum Technologies Center, Kutuzovski prospect 32, Moscow, 121170, Russia}

\date{October, 2026}

\begin{abstract}
Fusion gates enable the creation of large entangled photonic states for linear-optical quantum computing and communication protocols. However, their success probability is limited to $1/2$ without additional photonic resources. Previous schemes for boosting type-I fusion have used four ancillary single photons or a two-qubit Bell state, whose preparation is probabilistic and resource intensive.
Here we propose a boosted type-I fusion gate using only two independent ancillary single photons and no entangled ancillary input. It achieves a direct success probability of $5/8$, matching that of the previous four-single-photon proposal while using half as many ancillary photons.
The circuit is predominantly passive, with a single active element that applies a phase shift or swaps two modes according to intermediate photodetection outcomes.
Unlike the previous schemes, successful outputs occupy fixed modes and require no further conditional output routing. In particular, we show that the gate is more robust to heralded ancillary-photon loss than previously proposed boosted gates and reduces the average number of single photons required to generate a six-qubit ring graph state by more than $4.6\%$ compared with other type-I schemes. It therefore offers a resource-efficient primitive for scalable fusion-based photonic quantum information processing.
\end{abstract}

\maketitle

\section{Introduction}

Linear optics offers a promising platform for quantum computing and quantum communication~\cite{Knill2001,Browne2005,Kok2007,OBrien2009}, and recent advances in integrated photonic technologies provide a technological foundation for it's scalable implementation~\cite{Carolan2015,Wang2019,Kondratyev2026,Maring2024,PsiQuantum2025,AghaeeRad2025}. Weak interactions between photons, however, make entangling operations probabilistic on this platform. Modern approaches, such as measurement-based and fusion-based quantum computing, overcome this limitation by using probabilistic entangling measurements known as fusion gates~\cite{Browne2005}. In measurement-based quantum computing, fusion gates construct large entangled resource states on which computation is subsequently performed~\cite{GimenoSegovia2015, Pant2019}. In fusion-based quantum computing, computation is performed by applying fusion operations to a set of pre-generated entangled resource states~\cite{Omkar2022, Bartolucci2023}. These states can, in turn, be prepared by fusing even smaller seed states~\cite{Bartolucci2021, Pankovich2024_flexible, Staudacher2026, Melkozerov2026fusion}.

The success probabilities and physical complexity of fusion gates directly affect photonic resource requirements and tolerable errors in scalable linear-optical quantum computing architectures~\cite{Bartolucci2023, Melkozerov2024, Chan2026}. Measurement-based and fusion-based approaches can tolerate nondeterministic fusion gates provided that their success probabilities exceed the thresholds for the corresponding networks. For example, representative lossless thresholds are approximately $0.57$ for a $(2,2)$-Shor-encoded fusion network~\cite{Bartolucci2023} and $0.611$ for a four-dimensional 3-GHZ-based percolation lattice~\cite{Pant2019}.

In dual-rail encoding, standard type-I and type-II fusion gates have a maximum success probability of $1/2$ without additional ancillary photonic states~\cite{Calsamiglia2001, Browne2005}. Type-II fusion implements a Bell-state measurement and destroys both input photonic qubits, whereas type-I fusion implements an entangling map and destroys only one. This makes type-I fusion attractive for resource-efficient preparation of photonic graph and GHZ states~\cite{Bartolucci2021, Pankovich2024_flexible, Staudacher2026, Melkozerov2026fusion}.

For type-II fusion, entangled and single-photon ancillary states can raise the success probability above the $1/2$ limit~\cite{Grice2011,Ewert2014,Bartolucci2021,Hauser2025, Bayerbach2023,Guo2026}. For type-I fusion, an ancillary Bell pair can raise the success probability to $3/4$~\cite{Bartolucci2021}. A more recent scheme instead uses four independent ancillary single photons, avoiding entangled ancillary inputs and yielding a direct type-I fusion probability of $5/8$; distillation increases this probability to $11/16$ after one stage and asymptotically to $3/4$~\cite{Melkozerov2026fusion}. Here we introduce a resource-efficient type-I fusion gate that reaches a success probability of $5/8$ using only two independent ancillary single photons. A single fast active element conditionally applies the identity, a single-mode phase shift, or a swap of two fixed modes according to photodetection outcomes in a subset of the measured modes; all other transformations are passive. The gate produces event-ready outputs in fixed modes and requires no further conditional output routing, unlike previous proposals~\cite{Bartolucci2021, Melkozerov2026fusion}.

The construction avoids the overhead of preparing ancillary photonic Bell states, whose generation requires at least four single-photon inputs in a probabilistic linear-optical circuit~\cite{Stanisic2017}. Compared with the previous single-photon construction, it uses half as many ancillary photons while retaining the same direct fusion probability of $5/8 = 0.625$~\cite{Melkozerov2026fusion}. This probability exceeds the cited lossless thresholds for representative measurement-based and fusion-based quantum computing architectures~\cite{Bartolucci2023, Pant2019}.

Fusion gates also enable the preparation of entangled resource states for linear-optical quantum communication and computing protocols~\cite{Bartolucci2021, Wein2025, Melkozerov2026fusion}. Practical realization of these protocols requires reducing both photon cost and the number of active and passive optical elements used in resource-state preparation~\cite{Switch_networks2021, Bartolucci2021, Melkozerov2024, Chan2026}. In the scenarios considered, we show that the proposed gate improves error tolerance and reduces the photonic resources required to generate large entangled states, addressing a key bottleneck in scalable optical quantum computing.

The paper is organized as follows. Section~\ref{sec:standard_typeI} reviews standard type-I fusion and introduces the notation. Section~\ref{sec:boosting} presents the proposed circuit and describes its operation. Section~\ref{sec:performance} compares the gate with previous proposals and examines scenarios in which it improves error tolerance and reduces photonic resource requirements. Finally, Sec.~\ref{sec:conclusion} discusses the practical significance of the scheme and prospects for further applications.

\section{Type-I fusion}\label{sec:standard_typeI}

In dual-rail encoding, a logical qubit is represented by a single photon occupying one of two optical modes:
$
|0\rangle = |10\rrangle = \hat{a}_1^{\dagger}|\text{vac}\rrangle$ and $
|1\rangle = |01\rrangle = \hat{a}_2^{\dagger}|\text{vac}\rrangle.
$
Here $\hat{a}_1^{\dagger}$ and $\hat{a}_2^{\dagger}$ are creation operators, and $|\text{vac}\rrangle$ denotes the vacuum state. We use double kets for explicit optical Fock states and single kets for state labels and logical qubit states.

\begin{figure}[t]
	\includegraphics[width=0.45\textwidth]{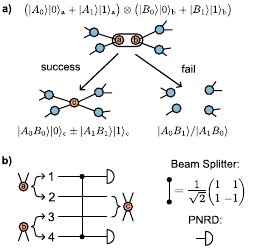}
	\caption{\textbf{Creation of multiqubit entangled states using fusion gates}: (a) A fusion gate combines the input states into a larger entangled state with success probability $P_s$. (b) One possible optical circuit for the standard type-I fusion gate, which achieves the maximum success probability of $P_s=1/2$ without ancillary photonic resources. Success is heralded by detecting exactly one photon in total across the two PNRDs; all other detection patterns indicate failure.}
	\label{fig:StandardType1}
\end{figure}

Type-I fusion was initially introduced to probabilistically connect separate entangled states \cite{Browne2005}, as shown in Fig.~\ref{fig:StandardType1}(a). Consider two maximally entangled states of $N_A$ and $N_B$ qubits, respectively. Their joint state has the form $\bigl( |A_0\rangle |0\rangle_a + |A_1\rangle |1\rangle_a \bigr)/\sqrt{2} \otimes  \bigl(|B_0\rangle |0\rangle_b+ |B_1\rangle |1\rangle_b \bigr)/\sqrt{2},$ where $\langle A_i|A_j\rangle = \langle B_i|B_j\rangle = \delta_{ij}$. These can be arbitrary graph states, Bell states, or multiqubit GHZ states, the most relevant states for linear-optical quantum computing and communication~\cite{Melkozerov2026fusion}. If modes $1$--$2$ encode qubit $a$ and modes $3$--$4$ encode qubit $b$, the corresponding Fock-state form is
\begin{equation}\label{eq:in_state_fock}
	\begin{aligned}
		| \text{in}\rangle = \frac{1}{2}\bigl(|A_0\rangle \hat{a}^{\dagger}_1 + |A_1\rangle \hat{a}^{\dagger}_2 \bigr) \otimes \bigl(|B_0\rangle \hat{a}^{\dagger}_3 + |B_1\rangle \hat{a}^{\dagger}_4 \bigr)|\text{vac}\rrangle
	\end{aligned}
\end{equation}

Figure~\ref{fig:StandardType1}(b) shows one possible linear-optical circuit for standard type-I fusion. Balanced $50:50$ beam splitters are depicted as dots connected by vertical lines between the corresponding modes. The circuit transforms the creation operators of the input optical state according to $a_i^{\dagger} \rightarrow \sum_{j=1}^{m} U_{ji} a_j^{\dagger}$, where $U$ is the unitary matrix describing the linear-optical transformation. Thus, when qubits $a$ and $b$ of the initial state \eqref{eq:in_state_fock} enter the circuit in Fig.~\ref{fig:StandardType1}(b), the state before measurement is

\begin{equation}\label{eq:type_1_res}
	\begin{aligned}
		\hat{U}_{fus}&|\text{in}\rangle = \Bigl(|{A}_0 {B}_1\rangle\frac{ (\hat{a}^{\dagger}_1)^2 -(\hat{a}^{\dagger}_4)^2 }{4}  + |A_1 B_0\rangle\frac{\hat{a}^{\dagger}_2 \hat{a}^{\dagger}_3}{2}  +   \\
		&+|A_0 B_0\rangle\frac{\hat{a}^{\dagger}_1 + \hat{a}^{\dagger}_4}{2\sqrt{2}} \hat{a}^{\dagger}_3  +|A_1 B_1\rangle\frac{\hat{a}^{\dagger}_1 - \hat{a}^{\dagger}_4}{2\sqrt{2}} \hat{a}^{\dagger}_2 \Bigr) |\text{vac}\rrangle.
	\end{aligned}
\end{equation}

The action of photon-number-resolving detectors (PNRDs) on the optical state can be represented by Kraus operators of the form $K_{\boldsymbol{n}} = \bigotimes_{i=1}^{m}|0\rrangle\llangle n_i| = |\boldsymbol{0}\rrangle\llangle \boldsymbol{n}|$,
where $n_i$ is the number of photons registered by the PNRD in mode $i$, and $\boldsymbol{n}=(n_1,\ldots,n_m)$ is the detection pattern in modes $\{1, \ldots, m\}$. Photon detection destroys photons in the measured modes, so we discard those modes after measurement.

If the detectors register exactly one photon in total in modes $\{1,4\}$ in Fig.~\ref{fig:StandardType1}(b), the corresponding Kraus operator is $K_{(1,0)}=|0_1 0_4\rrangle \llangle 1_1 0_4|$ or $K_{(0,1)}=|0_1 0_4\rrangle \llangle 0_1 1_4|$. Applying it to the state after the circuit gives
\begin{equation}\label{eq:type_1_suc}
	\begin{aligned}
		&|\text{out}\rangle = \bigl(K_{\boldsymbol{n}} \: \hat{U}_{fus}|\text{in}\rangle \bigr)/ \sqrt{P_{\boldsymbol{n}}} =\frac{1}{\sqrt{2}}\bigl( |A_0 B_0\rangle \hat{a}^{\dagger}_3 \pm \\
		&\pm  |A_1 B_1\rangle \hat{a}^{\dagger}_2\bigr) |\text{vac}\rrangle =\frac{1}{\sqrt{2}}\bigl( |A_0 B_0\rangle |0\rangle_c \pm  |A_1 B_1\rangle|1\rangle_c\bigr),
	\end{aligned}
\end{equation}
where the sign $\pm$ depends on the detection pattern and $P_{\boldsymbol{n}} = || K_{\boldsymbol{n}}  \hat{U}_{fus} |\text{in} \rangle||^2$ is the probability of the corresponding pattern $\boldsymbol{n}$.

The output is therefore an entangled state of $N_A + N_B - 1$ qubits, which corresponds to a successful fusion. Modes $\{2,3\}$ encode the surviving qubit $c$: $\hat{a}^{\dagger}_3 | \text{vac} \rrangle =|0_2 1_3\rrangle \equiv |0\rangle_c$ and $\hat{a}^{\dagger}_2 | \text{vac} \rrangle = |1_2 0_3\rrangle \equiv |1\rangle_c$. The total probability of these successful outcomes is
\begin{equation}\label{eq:type_1_prob}
	P_s = P_{(1,0)} + P_{(0,1)} = \frac{1}{2},
\end{equation}
which is the maximum fusion success probability without additional photonic resources \cite{Calsamiglia2001}.

Type-I fusion gate, therefore, implements the entangling map $|0\rangle\langle00| \pm |1\rangle\langle11|$ when successful. It destroys only one of the two input photons in contrast to the type-II fusion gate, which destroys them both. We also discuss how measuring only a subset of the input modes affects photon-loss detection in Appendix~\ref{app:loss_detection}.

When the detectors register either two photons or no photons across modes $\{1\}$ and $\{4\}$, which corresponds to $\boldsymbol{n} = (0,2), (2,0),$ or $(0,0)$, the fusion fails. The remaining states have the form $|A_0 B_1\rangle|0_2 0_3\rrangle$ for the two-photon outcomes and $|A_1 B_0\rangle |1_2 1_3\rrangle$ for the no-photon outcome.

\section{Boosting with two single photons}\label{sec:boosting}
\begin{figure}[t]
	\includegraphics[width=0.43\textwidth]{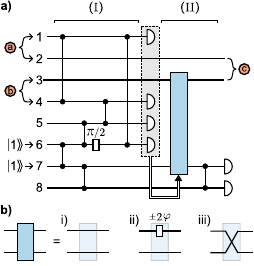}
	\caption{\textbf{The proposed linear-optical type-I fusion gate boosted by two single photons}. (a) Optical circuit of the gate. (b) The only active element applies (i) the identity operation, (ii) a phase shift on mode $\{3\}$, or (iii) a swap of modes $\{3\}$ and $\{7\}$, depending on the measurement outcomes in modes $\{1,4-6\}$. In practice, this element can readily be implemented with a single Mach-Zehnder interferometer.}
	\label{fig:BoostedType1}
\end{figure}
The idea behind the boosting scheme is to introduce an appropriate ancillary state that allows the terms in Eq.~\eqref{eq:type_1_res} with zero or two photons in modes $1$ and $4$ to contribute to the same detection patterns in the measured modes. The gate in Fig.~\ref{fig:BoostedType1} uses the state $|\text{anc}\rangle = |0_5 1_6 1_7 0_8\rrangle$ of two ancillary single photons in four optical modes $\{5-8\}$.

Qubits $a$ and $b$ of the initial state \eqref{eq:in_state_fock} enter modes $\{1,2\}$ and $\{3,4\}$ of the gate in Fig.~\ref{fig:BoostedType1}, respectively, together with the ancillary state $|\text{anc}\rangle$.

First, the static part $(I)$ of the circuit transforms the total input state into $\hat{U}_{(I)}\bigl(|\text{in}\rangle\otimes|0_5 1_6 1_7 0_8\rrangle\bigr)$. PNRDs then measure modes $\{1,4,5,6\}$. For a measurement pattern $\boldsymbol{n}=(n_1,n_4,n_5,n_6)$, the intermediate output state is
\begin{equation}\label{eq:intermediate_results}
	|\text{out}^{(\boldsymbol{n})}\rangle  = \left(K_{\boldsymbol{n}} \: \hat{U}_{(I)}(|\text{in}\rangle\otimes|0_5 1_6 1_7 0_8\rrangle)\right)/\sqrt{P_{\boldsymbol{n}}},
\end{equation}
where the probability of obtaining the corresponding pattern is
\begin{equation}\label{eq:pattern_prob}
	P_{\boldsymbol{n}} = ||K_{\boldsymbol{n}} \hat{U}_{(I)} (|\text{in}\rangle\otimes|0_5 1_6 1_7 0_8\rrangle) ||^2.
\end{equation}

We classify the states in \eqref{eq:intermediate_results} by the total number of detected photons and their distribution between the mode pairs $\{1,6\}$ and $\{4,5\}$. The individual measurement patterns are listed in Appendix~\ref{sec:app_outcome_classes}. The four classes are:

\begin{enumerate}[label=\roman*.]
	\item All single-photon events and three-photon events with either all photons in one mode pair or three singly occupied modes yield
	\begin{equation}\label{eq:out_i}
		\begin{aligned}
			|\text{out}^{(\text{i})}\rangle  &= \frac{1}{\sqrt{2}} \bigl( |A_0B_0 \rangle |0_21_3\rrangle \pm \\
			&\pm|A_1B_1 \rangle |1_20_3\rrangle \bigr) \otimes | Y_{7,8} \rangle,
		\end{aligned}
	\end{equation}
	where $|Y_{7,8}\rangle$ is a known by-product in modes $7,8$: $|Y\rangle = |00\rrangle$ when three photons are detected in total, and $|Y\rangle = | Y^{(2)} \rangle \equiv (|20\rrangle - \sqrt{2}|11\rrangle + |02\rrangle)/2$ when one photon is detected. The total probability of the patterns in class (i), $\{ \boldsymbol{n}_k \}_{(i)}$, is
	\begin{equation}\label{eq:out_i_prob}
		P^{(\text{i})}  = \frac{11}{32}.
	\end{equation}
	\item Three-photon events with a $(2+1)$ distribution between the mode pairs, where the two photons are bunched in one mode, yield
	\begin{equation}\label{eq:out_ii}
		\begin{aligned}
			|\text{out}^{(\text{ii})}\rangle  &= \frac{1}{\sqrt{2}} \bigl( e^{\pm i\varphi}|A_0B_0 \rangle |0_21_3\rrangle \pm\\
			&\pm e^{\mp i\varphi}|A_1B_1 \rangle |1_20_3\rrangle \bigr) \otimes | 0_7 0_8 \rrangle,
		\end{aligned}
	\end{equation}
	where $\varphi = \arctan{(2)}$. The total probability of the patterns in this class is
	\begin{equation}
		P^{(\text{ii})} = \frac{5}{32}.
	\end{equation}
	\item Two-photon events with both photons in the same mode pair yield
	\begin{equation}\label{eq:out_iii}
		\begin{aligned}
			|\text{out}^{(\text{iii})}\rangle &= \frac{1}{\sqrt{3}} \bigl(\sqrt{2}  |A_0B_1 \rangle |0_20_3\rrangle \otimes | Y^{(2)}_{7,8} \rangle \pm \\
			&\pm |A_1B_0 \rangle |1_21_3\rrangle \otimes | 0_7 0_8 \rrangle \bigr).
		\end{aligned}
	\end{equation}
	The total probability of the patterns in this class is
	\begin{equation}
		P^{(\text{iii})} = \frac{3}{16}.
	\end{equation}
	\item Vacuum events, two-photon events with one photon in each mode pair, and all four-photon events yield
	\begin{equation}\label{eq:out_iv}
		\begin{aligned}
			|\text{out}^{(\text{iv})}\rangle  &= |A_1 B_0\rangle |1_21_3\rrangle\otimes| Y_{7,8} \rangle, \text{ or}\\
			&=|A_0 B_1\rangle |0_2 0_3\rrangle \otimes |0_70_8 \rrangle,
		\end{aligned}
	\end{equation}
	where $|Y\rangle = |00\rrangle$ when two photons are detected in total and $|Y\rangle = | Y^{(2)}\rangle$ when none are detected. The total probability of the patterns in this class is
	\begin{equation}
		P^{(\text{iv})} = \frac{5}{16}.
	\end{equation}
\end{enumerate}

Patterns in class (i) directly produce successful outcomes of the standard fusion gate \eqref{eq:type_1_suc} and require no further transformation.

Outputs in class (ii) differ from standard successful outcomes by a measurement-dependent relative phase. In regime (ii), the active element in Fig.~\ref{fig:BoostedType1}(b) corrects this phase by applying the phase factor $e^{\mp 2 i \varphi}$ to mode $\{3\}$, with the sign determined by the measurement pattern $\boldsymbol{n}$.

For both classes (i) and (ii), the known by-product states in modes $\{7,8\}$ are discarded. Measuring these modes in the gate of Fig.~\ref{fig:BoostedType1} imposes no additional postselection.

For class (iii), the active element in Fig.~\ref{fig:BoostedType1}(b) swaps modes $\{3\}$ and $\{7\}$ of the state \eqref{eq:out_iii}. The final output then depends on the measurement pattern in modes $\{7,8\}$. If exactly one photon is detected in these modes, which occurs with conditional probability $P_1 = 2/3$ for class (iii), the resulting output is
\begin{equation}\label{eq:out_iii_s}
	\begin{aligned}
		|\text{out}^{(\text{iii})}_s\rangle  &= \frac{1}{\sqrt{2}} \bigl(  |A_0B_1 \rangle |0_2 1_3\rrangle \pm |A_1B_0 \rangle |1_2 0_3\rrangle\bigr).
	\end{aligned}
\end{equation}

For arbitrary Bell, GHZ, or graph states as inputs, this state is local Clifford equivalent to the state \eqref{eq:type_1_suc}; see, for example, Ref.~\cite{Melkozerov2026fusion}. We therefore count these events as successful. Their total probability is $P^{\text{(iii)}}_s = P^{\text{(iii)}} P_1 = 1/8$.

The total fusion success probability is the sum of the contributions from classes (i), (ii), and (iii):
\begin{equation}\label{eq:boosted_prob}
	P_{s} = P^{\text{(i)}} + P^{\text{(ii)}} + P^{\text{(iii)}}_s = \frac{5}{8}.
\end{equation}

For a class-(iii) outcome, detecting zero or two photons in modes $\{7,8\}$ instead heralds failure and leaves unentangled outputs of the form $|A_0B_1\rangle|0_22_3\rrangle$ or $|A_0B_1\rangle|0_20_3\rrangle$. Together with class (iv) outcomes, these events constitute the full set of failure outcomes of the boosted gate.

\section{Gate performance}\label{sec:performance}
\begin{figure*}[ht]
    \includegraphics[width=0.95\textwidth]{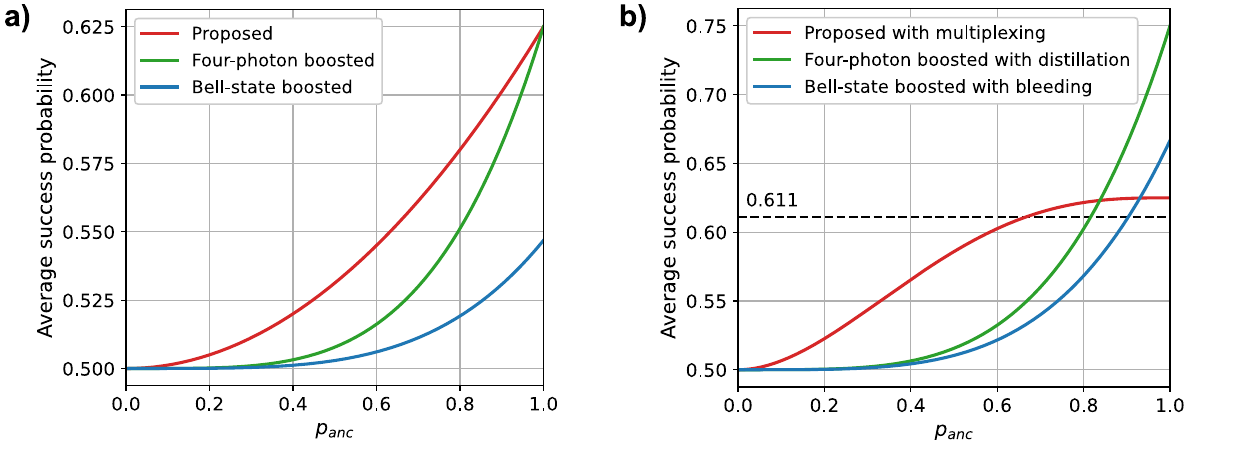}
    \caption{\textbf{(a) Average success probabilities of different boosted type-I fusion gates when ancillary single photons are generated probabilistically}. (b) Estimates for the proposed gate with $4 \rightarrow 2$ multiplexing and for the gates of Refs.~\cite{Bartolucci2021, Melkozerov2026fusion} in the $k \rightarrow \infty$ limit of the bleeding and distillation procedures, respectively. The dashed line indicates the fusion success threshold for a representative 3-GHZ percolation-based quantum computing architecture~\cite{Pant2019}.}
    \label{fig:Avg_prob}
\end{figure*}

\subsection{Robustness to probabilistic heralded ancilla preparation}\label{sec:prob_anc}
The proposed gate requires only two ancillary single photons for boosting, making it less sensitive to ancillary-state preparation failures than previously proposed boosting techniques~\cite{Bartolucci2021, Melkozerov2026fusion}.

In practice, single photons are generated probabilistically, so a complete ancillary state is not always available. If an ancillary state preparation failure is heralded, a boosted fusion gate can still operate with the standard success probability of $1/2$. Let $p$ be the probability that a complete ancillary state is generated. The average fusion success probability is then
\begin{equation}\label{eq:avg_succ_prob}
    P_{s}^{\text{avg}} = \left(P_s - \frac{1}{2}\right)p + \frac{1}{2}.
\end{equation}

Suppose each ancillary single photon is generated independently with heralded probability $p_{anc}$, as discussed in Appendix~\ref{sec:app_heralded_sps}. Then $p = p_{anc}^2$ for the proposed gate and $p = p_{anc}^4$ for the four-single-photon-boosted gate~\cite{Melkozerov2026fusion}. For the Bell-state-boosted gate~\cite{Bartolucci2021}, $p = 3p_{anc}^4/16$ if the Bell state is generated from a minimum of four single photons~\cite{Stanisic2017} with success probability $3/16$. This is the highest success probability among four-photon Bell-state generation schemes that do not require extensive use of active elements~\cite{Forbes2025}.

Figure~\ref{fig:Avg_prob}(a) shows the estimated average success probabilities of these three boosted gates. The proposed gate appears least sensitive to failures in ancillary-state preparation over the full range of $p_{anc}$.

The gates of Refs.~\cite{Melkozerov2026fusion, Bartolucci2021} can also benefit from distillation and bleeding procedures. Bleeding increases the probability of generating a Bell state from four single photons to as much as $2/3$~\cite{Bartolucci2021}. A closely related distillation procedure allows the four-single-photon-boosted gate to reach a success probability of up to $3/4$~\cite{Melkozerov2026fusion}. Both procedures, however, require extensive use of additional active and passive optical elements.

We compare these modifications with a simple modification of the proposed gate. Suppose four single photons are generated independently, matching the minimum number required to boost the other type-I fusion gates~\cite{Bartolucci2021, Melkozerov2026fusion}. A $4 \rightarrow 2$ multiplexing scheme then routes two successfully generated photons to the proposed gate. The probability that at least two photons are generated is
\begin{equation}
    p = 3p_{anc}^4 - 8p_{anc}^3+ 6p_{anc}^2.
\end{equation}

Figure~\ref{fig:Avg_prob}(b) shows the corresponding average success probabilities under probabilistic ancillary-photon generation. Even if the distillation and bleeding schemes are ideal and lossless, the proposed gate has higher tolerance to ancillary photon preparation failures for $p_{\text{anc}} < 0.84$. Moreover, it uses the same number of ancillary single-photon sources and a substantially simpler optical scheme.

Advanced linear-optical quantum computing architectures require fusion success probabilities to exceed particular thresholds for error correction~\cite{Bartolucci2023, GimenoSegovia2015, Pant2019}. For example, a four-dimensional percolation scheme based on three-photon GHZ states requires $P_s > 0.611$~\cite{Pant2019}. The proposed gate's average success probability crosses this threshold at approximately $p_{\text{anc}} = 0.67$. The corresponding values are approximately $0.82$ for the $3/4$-efficient four-photon-boosted gate~\cite{Melkozerov2026fusion} and $0.90$ for the Bell-state-boosted gate with bleeding~\cite{Bartolucci2021}, as shown in Fig.~\ref{fig:Avg_prob}(b).

\subsection{Sequential linear cluster state generation}
\begin{table}[b]
\caption{\textbf{Average number of ancillary single photons} required to add one qubit to a linear cluster by fusing it with Bell states.}
\label{tab:linear_cluster_growth}
\begin{ruledtabular}
\begin{tabular}{cccc}
& \begin{tabular}[c]{@{}c@{}} Bell-state- \\ boosted gate \cite{Bartolucci2021}\end{tabular} 
& \begin{tabular}[c]{@{}c@{}} Four-photon- \\ boosted gate \cite{Melkozerov2026fusion}\end{tabular} 
& \begin{tabular}[c]{@{}c@{}} \textbf{Proposed}\\ \textbf{gate} \end{tabular}  \\
[2pt] \hline
$n^{avg}_{anc}$ & 42.7 & 16.0 & \textbf{8.0} 

\end{tabular}
\end{ruledtabular}
\end{table}
Fusion gates can generate large entangled states by sequentially fusing smaller states~\cite{Bartolucci2021,Browne2005,GimenoSegovia2015}. Type-I gates are particularly useful for this task because they destroy only one of the two input photons, whereas type-II gates destroy both. This reduces the photonic resources needed to prepare large entangled resource states by fusion~\cite{Melkozerov2026fusion}.

If Bell states can be generated sequentially~\cite{Barz2010,Gubarev2020,Fldzhyan2021,Forbes2025,Skryabin2025}, type-I fusion gates can join them into arbitrarily large linear cluster states, which can serve, for example, as resources for measurement-based quantum computing~\cite{Browne2005} and quantum communication~\cite{Bell2014, deJong2023}. Bell states are local-Clifford equivalent to two-qubit linear cluster states, and a single type-I fusion can join two of them into a three-qubit linear cluster. The cluster can then be extended by fusing it with further two-qubit states, one at a time.

A standard type-I gate adds one qubit to the cluster when it succeeds and removes one when it fails. The average change in cluster size per fusion attempt with a two-qubit state is therefore $\Delta n =  P_s - (1-P_s) = 2P_s-1$. For standard fusion, $P_s = 1/2$ and $\Delta n = 0$. Boosting gives $\Delta n > 0$, allowing the linear cluster to grow on average.

The proposed gate adds $\Delta n = 2P_s-1 = 1/4$ qubits per attempt on average while using only two ancillary single photons. The average number of ancillary single photons required to add one qubit to the cluster is thus $n_{anc}^{avg} = n_{anc}/\Delta n = 8$.

The boosted gates of Refs.~\cite{Bartolucci2021, Melkozerov2026fusion} also have outcomes that preserve two qubits. Such outcomes, however, add one encoded qubit to a linear cluster rather than two qubits because of the entanglement structure of cluster states. For example, fusing two Bell states can produce a four-qubit GHZ state, which is not local-unitary equivalent to a four-qubit linear cluster~\cite{Raussendorf2003}. We therefore assume that the extra photons are measured in the Pauli-$X$ basis to obtain an ordinary linear cluster. For these gates, $\Delta n = P_s^{\Sigma} - (1-P_s^{\Sigma})= 2P_s^{\Sigma} - 1$, where $P_s^{\Sigma}$ is the total success probability. The gate of Ref.~\cite{Bartolucci2021} gives $\Delta n = 1/2$, and the $5/8$-efficient version of the gate of Ref.~\cite{Melkozerov2026fusion} gives $\Delta n = 1/4$.

Although the gate of Ref.~\cite{Bartolucci2021} provides a higher $\Delta n$, it requires more ancillary single photons for boosting. This increases the average number of ancillary single photons needed to add one qubit to the cluster, as shown in Table~\ref{tab:linear_cluster_growth}. Here we assume that the entangled two-qubit ancillary states used by the gate of Ref.~\cite{Bartolucci2021} are generated from four single photons with a success probability of $3/16$ in a perfectly resource-efficient multiplexing scheme (see Sec.~\ref{sec:Multiplexing}). The gate can also be boosted using the same Bell states used for cluster growth, but doing so would halve $\Delta n$ for the scheme and require additional active routing.

\subsection{Multiplexed generation of entangled states}\label{sec:Multiplexing}

\begin{table}[t]
\caption{\textbf{Average number of input single photons} required to generate three-qubit GHZ and six-qubit ring states with various type-I fusion gates.}
\label{tab:resource_estimations}
\begin{ruledtabular}
\begin{tabular}{ccccc}
\begin{tabular}[c]{@{}c@{}} Target \\ state\end{tabular} & \begin{tabular}[c]{@{}c@{}} Non-boosted \\$P_s = 1/2$ \cite{Browne2005}\end{tabular} 
& \begin{tabular}[c]{@{}c@{}} Bell-state \\ $P_s = 3/4$ \cite{Bartolucci2021}\end{tabular} 
& \begin{tabular}[c]{@{}c@{}} Four-photon \\ $P_s = 5/8$ \cite{Melkozerov2026fusion}\end{tabular} 
& \begin{tabular}[c]{@{}c@{}} \textbf{Proposed}\\$P_s = 5/8$\end{tabular}  \\
[2pt] \hline
$|\text{3-GHZ}\rangle$ & 85.3 & 85.3 & 74.7 & \textbf{71.5} \\
$|\text{G}_6\rangle$ & 824.9 & 673.2 & 635.3 & \textbf{605.7} \\

\end{tabular}
\end{ruledtabular}
\end{table}

Large entangled resource states for optical quantum protocols are commonly prepared by first generating small entangled states from single photons in probabilistic circuits and then fusing them into larger states~\cite{Browne2005, GimenoSegovia2015, Bartolucci2021, Staudacher2026}. To compare entanglement-generation strategies, we assume perfectly resource-efficient multiplexing~\cite{Bartolucci2021}. At each stage of a sequential protocol, a large number of identical probabilistic operations run in parallel, and all successful outputs are routed to the next stage. Under this assumption, the average number of input single photons required for a single successful output is $n_0/p_0$, where $n_0$ is the number of single photons consumed by one operation and $p_0$ is its success probability.

Table~\ref{tab:resource_estimations} shows the estimated photon requirements for generating two entangled resource states with different type-I fusion gates. We consider all-photonic fusion-based strategies for preparing three-qubit GHZ states $|\text{3-GHZ}\rangle$ and six-qubit ring states $|\text{G}_6\rangle$~\cite{Sahay2023}. The latter provide high robustness to losses and logical errors in fusion-based quantum computing architectures~\cite{Bartolucci2023}.

In all strategies, Bell states are first generated from four single photons each, with a success probability of $3/16$~\cite{Stanisic2017, Gubarev2020, Bartolucci2021, Forbes2025}. Two Bell states are then fused by the corresponding type-I gate to form a $|\text{3-GHZ}\rangle$ state. To create $|\text{G}_6\rangle$, three $|\text{3-GHZ}\rangle$ states are generated in this way. Two are then fused into a five-qubit GHZ state, which is fused with the third by two type-I gates to form the target $|\text{G}_6\rangle$~\cite{Sahay2023}. At this last stage, we use Bell-state-boosted gates~\cite{Bartolucci2021} in all strategies because they appear to be the most effective at this stage. We also assume a success probability of $5/8$ for the type-I gate of Ref.~\cite{Melkozerov2026fusion}, since its higher success probabilities require extensive use of active elements.

Three-qubit GHZ states can also be generated directly in probabilistic circuits~\cite{Forbes2025,Varnava2008,Gubarev2020,Bhatti2025,gimeno2016}. The highest success probability among known six-photon schemes is $1/32$~\cite{Varnava2008, Forbes2025}. Such a scheme, however, requires an average of $192$ single photons to generate one $|\text{3-GHZ}\rangle$ state.

These estimates show that the proposed gate reduces the average photon cost of preparing these entangled resource states even under the assumptions of perfectly resource-efficient multiplexing and ideal, lossless circuits.

\section{Conclusion}\label{sec:conclusion}
We have proposed an event-ready type-I photonic fusion gate that achieves a success probability of $5/8$ using two independent ancillary single photons. A single active element implements measurement-dependent phase corrections or a mode swap; all other transformations are passive. Successful outputs occupy a fixed pair of optical modes, allowing subsequent operations without conditional output routing. Compared with the four-single-photon construction of Ref.~\cite{Melkozerov2026fusion}, the gate halves the number of ancillary input photons while preserving the same direct fusion probability. It also avoids preparation of entangled ancillary state used in Ref.~\cite{Bartolucci2021}, whose probabilistic preparation requires at least four input single photons~\cite{Stanisic2017}.

We have evaluated the performance gate when ancillary single photons are supplied probabilistically by heralded sources. In this model, it appears to be more robust to ancillary-state preparation failures than previously proposed directly boosted gates~\cite{Melkozerov2026fusion, Bartolucci2021}. With $4\rightarrow2$ multiplexing, it also outperforms the idealized asymptotic distillation and bleeding benchmarks for single-photon generation probabilities below approximately $0.84$.

The reduced boosting overhead also benefits entangled-state preparation. When Bell pairs are supplied for sequential linear-cluster growth, the proposed gate requires an average of eight ancillary single photons per net added qubit, fewer than the other boosted gates under the ancillary-preparation assumptions considered here~\cite{Melkozerov2026fusion, Bartolucci2021}.

The gate also reduces the photonic cost of preparing three-qubit GHZ and six-qubit ring states, which serve as resource states for measurement-based and fusion-based quantum computing architectures~\cite{Pant2019, GimenoSegovia2015, Omkar2022, Bartolucci2023}. With perfectly resource-efficient multiplexing, the corresponding strategies require approximately $71$ and $606$ input single photons per state, respectively. These costs are $4.3\%$ and $4.7\%$ lower than those of the corresponding four-single-photon-boosted schemes.

To our knowledge, the specific four-mode state of two ancillary single photons used here has not previously been employed to boost a linear-optical entangling primitive. One possible extension is to adapt this boosting technique to multiqubit fusion~\cite{Pankovich2024_flexible, Pankovich2024, Patil2026}. Another is to investigate whether combining it with logical Bell-state measurements on encoded input qubits~\cite{Lee2015,Ewert2016,Kumar2026} could further reduce the photonic resources required to prepare large entangled states.

Together, these results make the gate a promising building block for resource-state preparation in scalable, error-tolerant linear-optical quantum computing and quantum communication protocols.

\section{Acknowledgments}
The authors’ work was supported by Rosatom in the framework of the Roadmap for Quantum computing (Contract № 868/1759-D dated 3 October 2025 and Contract №11-2025/1 dated 14 November 2025). 

A. Melkozerov is a recipient of a scholarship from the Theoretical Physics and Mathematics Advancement Foundation (BASIS) (No. 24-2-2-5-1).

\onecolumngrid
\appendix

\section{Photon-loss detection}\label{app:loss_detection}
A type-I fusion gate measures only a subset of its input modes, so photon loss is not always detected at the fusion stage. If a detection pattern heralds success, however, one photon is expected across the two modes encoding the surviving qubit. If later measurements detect no photon in either mode, they reveal a loss that was not detected during fusion.

This form of delayed loss detection is relevant, for example, when type-I fusion is used to prepare entangled resource states for measurement-based quantum computing~\cite{GimenoSegovia2015} or fusion-based quantum computing~\cite{Bartolucci2023}. In these protocols, the resource states are fully measured during computation, so losses that were not detected at the fusion stage can be identified and accounted for alongside other loss channels.

\section{Explicit outcome classes}\label{sec:app_outcome_classes}
Here we give the intermediate output states for the measurement patterns in each outcome class defined in Sec.~\ref{sec:boosting}. The detection pattern $\boldsymbol{n} = (n_1,n_4,n_5,n_6)$ gives the photon counts in modes $1$, $4$, $5$, and $6$, respectively.

\begin{enumerate}[label=\roman*.]
    \item
        \begin{enumerate}
            \item $\boldsymbol{n} = (1,0,0,0)/(0,1,0,0)/(0,0,1,0)/(0,0,0,1)$:
            \begin{equation}
                 |\text{out}^{(\text{i.a})}\rangle  = \frac{1}{\sqrt{2}} \bigl( |A_0B_0 \rangle |0_21_3\rrangle \pm|A_1B_1 \rangle |1_20_3\rrangle \bigr) \otimes | Y^{(2)}_{7,8} \rangle,
            \end{equation}
            where $| Y^{(2)}_{7,8} \rangle = (|20\rrangle - \sqrt{2}|11\rrangle + |02\rrangle)/2$. These outcomes have total probability $P^{(\text{i.a})} = 1/4$.

            \item $\boldsymbol{n} = (1,1,1,0)/(1,1,0,1)/(1,0,1,1)/(0,1,1,1)$, as well as all $\boldsymbol{n} = (n_1,n_4,n_5,n_6)$ for which $(n_1+n_6 = 3, n_4 + n_5 = 0)$ or $(n_1+n_6 = 0, n_4 + n_5 = 3)$:
            \begin{equation}
                 |\text{out}^{(\text{i.b})}\rangle  = \frac{1}{\sqrt{2}} \bigl( |A_0B_0 \rangle |0_21_3\rrangle\pm|A_1B_1 \rangle |1_20_3\rrangle \bigr) \otimes | 0_7 0_8 \rrangle.
            \end{equation}
            These outcomes have total probability $P^{(\text{i.b})} = 3/32$.
        \end{enumerate}

    \item
        \begin{enumerate}
            \item $\boldsymbol{n} = (2,1,0,0)/(0,0,1,2)/(0,2,0,1)/(1,0,2,0)$:
            \begin{equation}
                 |\text{out}^{(\text{ii.a})}\rangle  = \frac{1}{\sqrt{2}} \bigl( e^{i\varphi}|A_0B_0 \rangle |0_21_3\rrangle \pm e^{-i\varphi} |A_1B_1 \rangle |1_20_3\rrangle \bigr) \otimes | 0_70_8 \rrangle,
            \end{equation}
            where $\varphi = \arctan{(2)}$. These outcomes have total probability $P^{(\text{ii.a})} = 5/64$.

            \item $\boldsymbol{n} = (2,0,1,0)/(0,1,0,2)/(1,2,0,0)/(0,0,2,1)$:
            \begin{equation}
                 |\text{out}^{(\text{ii.b})}\rangle  = \frac{1}{\sqrt{2}} \bigl( e^{-i\varphi}|A_0B_0 \rangle |0_21_3\rrangle \pm e^{i\varphi} |A_1B_1 \rangle |1_20_3\rrangle \bigr) \otimes | 0_70_8 \rrangle.
            \end{equation}
            These outcomes have total probability $P^{(\text{ii.b})} = 5/64$.
        \end{enumerate}

    \item $\boldsymbol{n} = (2,0,0,0)/(1,0,0,1)/(0,0,0,2)/(0,2,0,0)/(0,1,1,0)/(0,0,2,0)$:
    \begin{equation}
         |\text{out}^{(\text{iii})}\rangle  = \frac{1}{\sqrt{3}} \bigl(\sqrt{2}  |A_0B_1 \rangle |0_20_3\rrangle \otimes | Y^{(2)}_{7,8} \rangle \pm |A_1B_0 \rangle |1_21_3\rrangle \otimes | 0_7 0_8 \rrangle \bigr).
    \end{equation}
    These outcomes have total probability$ P^{(\text{iii})} = 3/16$.

    \item
        \begin{enumerate}
            \item All $\boldsymbol{n} = (n_1,n_4,n_5,n_6)$ for which $(n_1+n_4+n_5+n_6 = 4)$:
            \begin{equation}
                |\text{out}^{(\text{iv.a})}\rangle  = |A_0 B_1\rangle |0_20_30_70_8\rrangle.
            \end{equation}
            These outcomes have total probability $P^{(\text{iv.a})} = 1/8.$

            \item $\boldsymbol{n} = (0,0,0,0)$:
            \begin{equation}
                |\text{out}^{(\text{iv.b})}\rangle  = |A_1 B_0\rangle |1_21_3\rrangle\otimes|Y^{(2)}_{7,8}\rangle.
            \end{equation}
            These outcomes have total probability $P^{(\text{iv.b})} = 1/8$.

            \item $\boldsymbol{n} = (1,1,0,0)/(1,0,1,0)/(0,1,0,1)/(0,0,1,1)$:
            \begin{equation}
                |\text{out}^{(\text{iv.c})}\rangle  = |A_1 B_0\rangle |1_21_3\rrangle\otimes|0_7 0_8\rrangle.
            \end{equation}
            These outcomes have total probability $P^{(\text{iv.c})} = 1/16$.
        \end{enumerate}
\end{enumerate}

\section{Heralded single-photon sources}\label{sec:app_heralded_sps}
The ancillary photons considered in Sec.~\ref{sec:prob_anc} can be supplied by several types of single-photon sources. One common approach uses spontaneous parametric down-conversion or spontaneous four-wave mixing to generate correlated photon pairs. Detection of one photon heralds the other \cite{Collins2013}. Increasing the pair-generation probability in either process also increases the rate of unwanted multipair events, creating a trade-off between the heralded-photon rate and multiphoton contamination.

Multiplexing can increase the probability of delivering a heralded single photon while keeping the generation probability of each source low. For example, a switching network can select and route a photon from one of several parallel heralded sources~\cite{Switch_networks2021, Kaneda2026}.

Another approach uses solid-state emitters such as quantum dots, which are often considered on-demand single-photon sources. Some quantum-dot emission processes, however, generate correlated photon pairs, allowing one photon to herald the other. This has been demonstrated experimentally~\cite{Liu2024}, suggesting that such emitters may also provide an efficient route to heralded single photons.


\begin{thebibliography}{49}
	\makeatletter
	\providecommand \@ifxundefined [1]{%
		\@ifx{#1\undefined}
	}%
	\providecommand \@ifnum [1]{%
		\ifnum #1\expandafter \@firstoftwo
		\else \expandafter \@secondoftwo
		\fi
	}%
	\providecommand \@ifx [1]{%
		\ifx #1\expandafter \@firstoftwo
		\else \expandafter \@secondoftwo
		\fi
	}%
	\providecommand \natexlab [1]{#1}%
	\providecommand \enquote  [1]{``#1''}%
	\providecommand \bibnamefont  [1]{#1}%
	\providecommand \bibfnamefont [1]{#1}%
	\providecommand \citenamefont [1]{#1}%
	\providecommand \href@noop [0]{\@secondoftwo}%
	\providecommand \href [0]{\begingroup \@sanitize@url \@href}%
	\providecommand \@href[1]{\@@startlink{#1}\@@href}%
	\providecommand \@@href[1]{\endgroup#1\@@endlink}%
	\providecommand \@sanitize@url [0]{\catcode `\\12\catcode `\$12\catcode
		`\&12\catcode `\#12\catcode `\^12\catcode `\_12\catcode `\%12\relax}%
	\providecommand \@@startlink[1]{}%
	\providecommand \@@endlink[0]{}%
	\providecommand \url  [0]{\begingroup\@sanitize@url \@url }%
	\providecommand \@url [1]{\endgroup\@href {#1}{\urlprefix }}%
	\providecommand \urlprefix  [0]{URL }%
	\providecommand \Eprint [0]{\href }%
	\providecommand \doibase [0]{https://doi.org/}%
	\providecommand \selectlanguage [0]{\@gobble}%
	\providecommand \bibinfo  [0]{\@secondoftwo}%
	\providecommand \bibfield  [0]{\@secondoftwo}%
	\providecommand \translation [1]{[#1]}%
	\providecommand \BibitemOpen [0]{}%
	\providecommand \bibitemStop [0]{}%
	\providecommand \bibitemNoStop [0]{.\EOS\space}%
	\providecommand \EOS [0]{\spacefactor3000\relax}%
	\providecommand \BibitemShut  [1]{\csname bibitem#1\endcsname}%
	\let\auto@bib@innerbib\@empty
	\bibitem [{\citenamefont {Knill}\ \emph {et~al.}(2001)\citenamefont {Knill},
		\citenamefont {Laflamme},\ and\ \citenamefont {Milburn}}]{Knill2001}%
	\BibitemOpen
	\bibfield  {author} {\bibinfo {author} {\bibfnamefont {E.}~\bibnamefont
			{Knill}}, \bibinfo {author} {\bibfnamefont {R.}~\bibnamefont {Laflamme}},\
		and\ \bibinfo {author} {\bibfnamefont {G.~J.}\ \bibnamefont {Milburn}},\
	}\bibfield  {title} {\bibinfo {title} {A scheme for efficient quantum
			computation with linear optics},\ }\href {https://doi.org/10.1038/35051009}
	{\bibfield  {journal} {\bibinfo  {journal} {Nature}\ }\textbf {\bibinfo
			{volume} {409}},\ \bibinfo {pages} {46–52} (\bibinfo {year}
		{2001})}\BibitemShut {NoStop}%
	\bibitem [{\citenamefont {Browne}\ and\ \citenamefont
		{Rudolph}(2005)}]{Browne2005}%
	\BibitemOpen
	\bibfield  {author} {\bibinfo {author} {\bibfnamefont {D.~E.}\ \bibnamefont
			{Browne}}\ and\ \bibinfo {author} {\bibfnamefont {T.}~\bibnamefont
			{Rudolph}},\ }\bibfield  {title} {\bibinfo {title} {Resource-efficient linear
			optical quantum computation},\ }\href
	{https://doi.org/10.1103/physrevlett.95.010501} {\bibfield  {journal}
		{\bibinfo  {journal} {Physical Review Letters}\ }\textbf {\bibinfo {volume}
			{95}},\ \bibinfo {pages} {1} (\bibinfo {year} {2005})}\BibitemShut {NoStop}%
	\bibitem [{\citenamefont {Kok}\ \emph {et~al.}(2007)\citenamefont {Kok},
		\citenamefont {Munro}, \citenamefont {Nemoto}, \citenamefont {Ralph},
		\citenamefont {Dowling},\ and\ \citenamefont {Milburn}}]{Kok2007}%
	\BibitemOpen
	\bibfield  {author} {\bibinfo {author} {\bibfnamefont {P.}~\bibnamefont
			{Kok}}, \bibinfo {author} {\bibfnamefont {W.~J.}\ \bibnamefont {Munro}},
		\bibinfo {author} {\bibfnamefont {K.}~\bibnamefont {Nemoto}}, \bibinfo
		{author} {\bibfnamefont {T.~C.}\ \bibnamefont {Ralph}}, \bibinfo {author}
		{\bibfnamefont {J.~P.}\ \bibnamefont {Dowling}},\ and\ \bibinfo {author}
		{\bibfnamefont {G.~J.}\ \bibnamefont {Milburn}},\ }\bibfield  {title}
	{\bibinfo {title} {Linear optical quantum computing with photonic qubits},\
	}\href {https://doi.org/10.1103/revmodphys.79.135} {\bibfield  {journal}
		{\bibinfo  {journal} {Reviews of Modern Physics}\ }\textbf {\bibinfo {volume}
			{79}},\ \bibinfo {pages} {135–174} (\bibinfo {year} {2007})}\BibitemShut
	{NoStop}%
	\bibitem [{\citenamefont {O’Brien}\ \emph {et~al.}(2009)\citenamefont
		{O’Brien}, \citenamefont {Furusawa},\ and\ \citenamefont
		{Vučković}}]{OBrien2009}%
	\BibitemOpen
	\bibfield  {author} {\bibinfo {author} {\bibfnamefont {J.~L.}\ \bibnamefont
			{O’Brien}}, \bibinfo {author} {\bibfnamefont {A.}~\bibnamefont
			{Furusawa}},\ and\ \bibinfo {author} {\bibfnamefont {J.}~\bibnamefont
			{Vučković}},\ }\bibfield  {title} {\bibinfo {title} {Photonic quantum
			technologies},\ }\href {https://doi.org/10.1038/nphoton.2009.229} {\bibfield
		{journal} {\bibinfo  {journal} {Nature Photonics}\ }\textbf {\bibinfo
			{volume} {3}},\ \bibinfo {pages} {687–695} (\bibinfo {year}
		{2009})}\BibitemShut {NoStop}%
	\bibitem [{\citenamefont {Carolan}\ \emph {et~al.}(2015)\citenamefont
		{Carolan}, \citenamefont {Harrold}, \citenamefont {Sparrow}, \citenamefont
		{Martín-López}, \citenamefont {Russell}, \citenamefont {Silverstone},
		\citenamefont {Shadbolt}, \citenamefont {Matsuda}, \citenamefont {Oguma},
		\citenamefont {Itoh} \emph {et~al.}}]{Carolan2015}%
	\BibitemOpen
	\bibfield  {author} {\bibinfo {author} {\bibfnamefont {J.}~\bibnamefont
			{Carolan}}, \bibinfo {author} {\bibfnamefont {C.}~\bibnamefont {Harrold}},
		\bibinfo {author} {\bibfnamefont {C.}~\bibnamefont {Sparrow}}, \bibinfo
		{author} {\bibfnamefont {E.}~\bibnamefont {Martín-López}}, \bibinfo
		{author} {\bibfnamefont {N.~J.}\ \bibnamefont {Russell}}, \bibinfo {author}
		{\bibfnamefont {J.~W.}\ \bibnamefont {Silverstone}}, \bibinfo {author}
		{\bibfnamefont {P.~J.}\ \bibnamefont {Shadbolt}}, \bibinfo {author}
		{\bibfnamefont {N.}~\bibnamefont {Matsuda}}, \bibinfo {author} {\bibfnamefont
			{M.}~\bibnamefont {Oguma}}, \bibinfo {author} {\bibfnamefont
			{M.}~\bibnamefont {Itoh}}, \emph {et~al.},\ }\bibfield  {title} {\bibinfo
		{title} {Universal linear optics},\ }\href
	{https://doi.org/10.1126/science.aab3642} {\bibfield  {journal} {\bibinfo
			{journal} {Science}\ }\textbf {\bibinfo {volume} {349}},\ \bibinfo {pages}
		{711–716} (\bibinfo {year} {2015})}\BibitemShut {NoStop}%
	\bibitem [{\citenamefont {Wang}\ \emph {et~al.}(2019)\citenamefont {Wang},
		\citenamefont {Sciarrino}, \citenamefont {Laing},\ and\ \citenamefont
		{Thompson}}]{Wang2019}%
	\BibitemOpen
	\bibfield  {author} {\bibinfo {author} {\bibfnamefont {J.}~\bibnamefont
			{Wang}}, \bibinfo {author} {\bibfnamefont {F.}~\bibnamefont {Sciarrino}},
		\bibinfo {author} {\bibfnamefont {A.}~\bibnamefont {Laing}},\ and\ \bibinfo
		{author} {\bibfnamefont {M.~G.}\ \bibnamefont {Thompson}},\ }\bibfield
	{title} {\bibinfo {title} {Integrated photonic quantum technologies},\ }\href
	{https://doi.org/10.1038/s41566-019-0532-1} {\bibfield  {journal} {\bibinfo
			{journal} {Nature Photonics}\ }\textbf {\bibinfo {volume} {14}},\ \bibinfo
		{pages} {273–284} (\bibinfo {year} {2019})}\BibitemShut {NoStop}%
	\bibitem [{\citenamefont {Kondratyev}\ \emph {et~al.}(2026)\citenamefont
		{Kondratyev}, \citenamefont {Urusova}, \citenamefont {Argenchiev},
		\citenamefont {Klushnikov}, \citenamefont {Kuzmin}, \citenamefont {Skryabin},
		\citenamefont {Golikov}, \citenamefont {Kovalyuk}, \citenamefont {Goltsman},
		\citenamefont {Dyakonov} \emph {et~al.}}]{Kondratyev2026}%
	\BibitemOpen
	\bibfield  {author} {\bibinfo {author} {\bibfnamefont {I.}~\bibnamefont
			{Kondratyev}}, \bibinfo {author} {\bibfnamefont {K.}~\bibnamefont {Urusova}},
		\bibinfo {author} {\bibfnamefont {A.}~\bibnamefont {Argenchiev}}, \bibinfo
		{author} {\bibfnamefont {N.}~\bibnamefont {Klushnikov}}, \bibinfo {author}
		{\bibfnamefont {S.}~\bibnamefont {Kuzmin}}, \bibinfo {author} {\bibfnamefont
			{N.}~\bibnamefont {Skryabin}}, \bibinfo {author} {\bibfnamefont
			{A.}~\bibnamefont {Golikov}}, \bibinfo {author} {\bibfnamefont
			{V.}~\bibnamefont {Kovalyuk}}, \bibinfo {author} {\bibfnamefont
			{G.}~\bibnamefont {Goltsman}}, \bibinfo {author} {\bibfnamefont
			{I.}~\bibnamefont {Dyakonov}}, \emph {et~al.},\ }\bibfield  {title} {\bibinfo
		{title} {Effective programming of a photonic processor with complex
			interferometric structure},\ }\href {http://dx.doi.org/10.1103/btbd-w8fk}
	{\bibfield  {journal} {\bibinfo  {journal} {Physical Review Applied}\
		}\textbf {\bibinfo {volume} {25}} (\bibinfo {year} {2026})}\BibitemShut
	{NoStop}%
	\bibitem [{\citenamefont {Maring}\ \emph {et~al.}(2024)\citenamefont {Maring},
		\citenamefont {Fyrillas}, \citenamefont {Pont}, \citenamefont {Ivanov},
		\citenamefont {Stepanov}, \citenamefont {Margaria}, \citenamefont {Hease},
		\citenamefont {Pishchagin}, \citenamefont {Lemaître}, \citenamefont {Sagnes}
		\emph {et~al.}}]{Maring2024}%
	\BibitemOpen
	\bibfield  {author} {\bibinfo {author} {\bibfnamefont {N.}~\bibnamefont
			{Maring}}, \bibinfo {author} {\bibfnamefont {A.}~\bibnamefont {Fyrillas}},
		\bibinfo {author} {\bibfnamefont {M.}~\bibnamefont {Pont}}, \bibinfo {author}
		{\bibfnamefont {E.}~\bibnamefont {Ivanov}}, \bibinfo {author} {\bibfnamefont
			{P.}~\bibnamefont {Stepanov}}, \bibinfo {author} {\bibfnamefont
			{N.}~\bibnamefont {Margaria}}, \bibinfo {author} {\bibfnamefont
			{W.}~\bibnamefont {Hease}}, \bibinfo {author} {\bibfnamefont
			{A.}~\bibnamefont {Pishchagin}}, \bibinfo {author} {\bibfnamefont
			{A.}~\bibnamefont {Lemaître}}, \bibinfo {author} {\bibfnamefont
			{I.}~\bibnamefont {Sagnes}}, \emph {et~al.},\ }\bibfield  {title} {\bibinfo
		{title} {A versatile single-photon-based quantum computing platform},\ }\href
	{https://doi.org/10.1038/s41566-024-01403-4} {\bibfield  {journal} {\bibinfo
			{journal} {Nature Photonics}\ }\textbf {\bibinfo {volume} {18}},\ \bibinfo
		{pages} {603–609} (\bibinfo {year} {2024})}\BibitemShut {NoStop}%
	\bibitem [{\citenamefont {Alexander}\ \emph {et~al.}(2025)\citenamefont
		{Alexander}, \citenamefont {Benyamini}, \citenamefont {Black}, \citenamefont
		{Bonneau}, \citenamefont {Burgos}, \citenamefont {Burridge}, \citenamefont
		{Cable}, \citenamefont {Campbell}, \citenamefont {Catalano}, \citenamefont
		{Ceballos} \emph {et~al.}}]{PsiQuantum2025}%
	\BibitemOpen
	\bibfield  {author} {\bibinfo {author} {\bibfnamefont {K.}~\bibnamefont
			{Alexander}}, \bibinfo {author} {\bibfnamefont {A.}~\bibnamefont
			{Benyamini}}, \bibinfo {author} {\bibfnamefont {D.}~\bibnamefont {Black}},
		\bibinfo {author} {\bibfnamefont {D.}~\bibnamefont {Bonneau}}, \bibinfo
		{author} {\bibfnamefont {S.}~\bibnamefont {Burgos}}, \bibinfo {author}
		{\bibfnamefont {B.}~\bibnamefont {Burridge}}, \bibinfo {author}
		{\bibfnamefont {H.}~\bibnamefont {Cable}}, \bibinfo {author} {\bibfnamefont
			{G.}~\bibnamefont {Campbell}}, \bibinfo {author} {\bibfnamefont
			{G.}~\bibnamefont {Catalano}}, \bibinfo {author} {\bibfnamefont
			{A.}~\bibnamefont {Ceballos}}, \emph {et~al.},\ }\bibfield  {title} {\bibinfo
		{title} {A manufacturable platform for photonic quantum computing},\ }\href
	{https://doi.org/10.1038/s41586-025-08820-7} {\bibfield  {journal} {\bibinfo
			{journal} {Nature}\ }\textbf {\bibinfo {volume} {641}},\ \bibinfo {pages}
		{876–883} (\bibinfo {year} {2025})}\BibitemShut {NoStop}%
	\bibitem [{\citenamefont {Aghaee~Rad}\ \emph {et~al.}(2025)\citenamefont
		{Aghaee~Rad}, \citenamefont {Ainsworth}, \citenamefont {Alexander},
		\citenamefont {Altieri}, \citenamefont {Askarani}, \citenamefont {Baby},
		\citenamefont {Banchi}, \citenamefont {Baragiola}, \citenamefont {Bourassa},
		\citenamefont {Chadwick} \emph {et~al.}}]{AghaeeRad2025}%
	\BibitemOpen
	\bibfield  {author} {\bibinfo {author} {\bibfnamefont {H.}~\bibnamefont
			{Aghaee~Rad}}, \bibinfo {author} {\bibfnamefont {T.}~\bibnamefont
			{Ainsworth}}, \bibinfo {author} {\bibfnamefont {R.~N.}\ \bibnamefont
			{Alexander}}, \bibinfo {author} {\bibfnamefont {B.}~\bibnamefont {Altieri}},
		\bibinfo {author} {\bibfnamefont {M.~F.}\ \bibnamefont {Askarani}}, \bibinfo
		{author} {\bibfnamefont {R.}~\bibnamefont {Baby}}, \bibinfo {author}
		{\bibfnamefont {L.}~\bibnamefont {Banchi}}, \bibinfo {author} {\bibfnamefont
			{B.~Q.}\ \bibnamefont {Baragiola}}, \bibinfo {author} {\bibfnamefont {J.~E.}\
			\bibnamefont {Bourassa}}, \bibinfo {author} {\bibfnamefont {R.~S.}\
			\bibnamefont {Chadwick}}, \emph {et~al.},\ }\bibfield  {title} {\bibinfo
		{title} {Scaling and networking a modular photonic quantum computer},\ }\href
	{https://doi.org/10.1038/s41586-024-08406-9} {\bibfield  {journal} {\bibinfo
			{journal} {Nature}\ }\textbf {\bibinfo {volume} {638}},\ \bibinfo {pages}
		{912–919} (\bibinfo {year} {2025})}\BibitemShut {NoStop}%
	\bibitem [{\citenamefont {Gimeno-Segovia}\ \emph {et~al.}(2015)\citenamefont
		{Gimeno-Segovia}, \citenamefont {Shadbolt}, \citenamefont {Browne},\ and\
		\citenamefont {Rudolph}}]{GimenoSegovia2015}%
	\BibitemOpen
	\bibfield  {author} {\bibinfo {author} {\bibfnamefont {M.}~\bibnamefont
			{Gimeno-Segovia}}, \bibinfo {author} {\bibfnamefont {P.}~\bibnamefont
			{Shadbolt}}, \bibinfo {author} {\bibfnamefont {D.~E.}\ \bibnamefont
			{Browne}},\ and\ \bibinfo {author} {\bibfnamefont {T.}~\bibnamefont
			{Rudolph}},\ }\bibfield  {title} {\bibinfo {title} {From three-photon
			greenberger-horne-zeilinger states to ballistic universal quantum
			computation},\ }\href {https://doi.org/10.1103/physrevlett.115.020502}
	{\bibfield  {journal} {\bibinfo  {journal} {Physical Review Letters}\
		}\textbf {\bibinfo {volume} {115}},\ \bibinfo {pages} {2} (\bibinfo {year}
		{2015})}\BibitemShut {NoStop}%
	\bibitem [{\citenamefont {Pant}\ \emph {et~al.}(2019)\citenamefont {Pant},
		\citenamefont {Towsley}, \citenamefont {Englund},\ and\ \citenamefont
		{Guha}}]{Pant2019}%
	\BibitemOpen
	\bibfield  {author} {\bibinfo {author} {\bibfnamefont {M.}~\bibnamefont
			{Pant}}, \bibinfo {author} {\bibfnamefont {D.}~\bibnamefont {Towsley}},
		\bibinfo {author} {\bibfnamefont {D.}~\bibnamefont {Englund}},\ and\ \bibinfo
		{author} {\bibfnamefont {S.}~\bibnamefont {Guha}},\ }\bibfield  {title}
	{\bibinfo {title} {Percolation thresholds for photonic quantum computing},\
	}\href {https://doi.org/10.1038/s41467-019-08948-x} {\bibfield  {journal}
		{\bibinfo  {journal} {Nature Communications}\ }\textbf {\bibinfo {volume}
			{10}},\ \bibinfo {pages} {1} (\bibinfo {year} {2019})}\BibitemShut {NoStop}%
	\bibitem [{\citenamefont {Omkar}\ \emph {et~al.}(2022)\citenamefont {Omkar},
		\citenamefont {Lee}, \citenamefont {Teo}, \citenamefont {Lee},\ and\
		\citenamefont {Jeong}}]{Omkar2022}%
	\BibitemOpen
	\bibfield  {author} {\bibinfo {author} {\bibfnamefont {S.}~\bibnamefont
			{Omkar}}, \bibinfo {author} {\bibfnamefont {S.-H.}\ \bibnamefont {Lee}},
		\bibinfo {author} {\bibfnamefont {Y.~S.}\ \bibnamefont {Teo}}, \bibinfo
		{author} {\bibfnamefont {S.-W.}\ \bibnamefont {Lee}},\ and\ \bibinfo {author}
		{\bibfnamefont {H.}~\bibnamefont {Jeong}},\ }\bibfield  {title} {\bibinfo
		{title} {All-photonic architecture for scalable quantum computing with
			greenberger-horne-zeilinger states},\ }\href
	{https://doi.org/10.1103/prxquantum.3.030309} {\bibfield  {journal} {\bibinfo
			{journal} {PRX Quantum}\ }\textbf {\bibinfo {volume} {3}},\ \bibinfo {pages}
		{3} (\bibinfo {year} {2022})}\BibitemShut {NoStop}%
	\bibitem [{\citenamefont {Bartolucci}\ \emph {et~al.}(2023)\citenamefont
		{Bartolucci}, \citenamefont {Birchall}, \citenamefont {Bombín},
		\citenamefont {Cable}, \citenamefont {Dawson}, \citenamefont
		{Gimeno-Segovia}, \citenamefont {Johnston}, \citenamefont {Kieling},
		\citenamefont {Nickerson}, \citenamefont {Pant} \emph
		{et~al.}}]{Bartolucci2023}%
	\BibitemOpen
	\bibfield  {author} {\bibinfo {author} {\bibfnamefont {S.}~\bibnamefont
			{Bartolucci}}, \bibinfo {author} {\bibfnamefont {P.}~\bibnamefont
			{Birchall}}, \bibinfo {author} {\bibfnamefont {H.}~\bibnamefont {Bombín}},
		\bibinfo {author} {\bibfnamefont {H.}~\bibnamefont {Cable}}, \bibinfo
		{author} {\bibfnamefont {C.}~\bibnamefont {Dawson}}, \bibinfo {author}
		{\bibfnamefont {M.}~\bibnamefont {Gimeno-Segovia}}, \bibinfo {author}
		{\bibfnamefont {E.}~\bibnamefont {Johnston}}, \bibinfo {author}
		{\bibfnamefont {K.}~\bibnamefont {Kieling}}, \bibinfo {author} {\bibfnamefont
			{N.}~\bibnamefont {Nickerson}}, \bibinfo {author} {\bibfnamefont
			{M.}~\bibnamefont {Pant}}, \emph {et~al.},\ }\bibfield  {title} {\bibinfo
		{title} {Fusion-based quantum computation},\ }\href
	{https://doi.org/10.1038/s41467-023-36493-1} {\bibfield  {journal} {\bibinfo
			{journal} {Nature Communications}\ }\textbf {\bibinfo {volume} {14}},\
		\bibinfo {pages} {1} (\bibinfo {year} {2023})}\BibitemShut {NoStop}%
	\bibitem [{\citenamefont {Bartolucci}\ \emph
		{et~al.}(2021{\natexlab{a}})\citenamefont {Bartolucci}, \citenamefont
		{Birchall}, \citenamefont {Gimeno-Segovia}, \citenamefont {Johnston},
		\citenamefont {Kieling}, \citenamefont {Pant}, \citenamefont {Rudolph},
		\citenamefont {Smith}, \citenamefont {Sparrow},\ and\ \citenamefont
		{Vidrighin}}]{Bartolucci2021}%
	\BibitemOpen
	\bibfield  {author} {\bibinfo {author} {\bibfnamefont {S.}~\bibnamefont
			{Bartolucci}}, \bibinfo {author} {\bibfnamefont {P.~M.}\ \bibnamefont
			{Birchall}}, \bibinfo {author} {\bibfnamefont {M.}~\bibnamefont
			{Gimeno-Segovia}}, \bibinfo {author} {\bibfnamefont {E.}~\bibnamefont
			{Johnston}}, \bibinfo {author} {\bibfnamefont {K.}~\bibnamefont {Kieling}},
		\bibinfo {author} {\bibfnamefont {M.}~\bibnamefont {Pant}}, \bibinfo {author}
		{\bibfnamefont {T.}~\bibnamefont {Rudolph}}, \bibinfo {author} {\bibfnamefont
			{J.}~\bibnamefont {Smith}}, \bibinfo {author} {\bibfnamefont
			{C.}~\bibnamefont {Sparrow}},\ and\ \bibinfo {author} {\bibfnamefont {M.~D.}\
			\bibnamefont {Vidrighin}},\ }\href@noop {} {\bibinfo {title} {Creation of
			entangled photonic states using linear optics}} (\bibinfo {year}
	{2021}{\natexlab{a}}),\ \Eprint {https://arxiv.org/abs/2106.13825}
	{arXiv:2106.13825} \BibitemShut {NoStop}%
	\bibitem [{\citenamefont {Pankovich}\ \emph
		{et~al.}(2024{\natexlab{a}})\citenamefont {Pankovich}, \citenamefont
		{Neville}, \citenamefont {Kan}, \citenamefont {Omkar}, \citenamefont {Wan},\
		and\ \citenamefont {Br\'adler}}]{Pankovich2024_flexible}%
	\BibitemOpen
	\bibfield  {author} {\bibinfo {author} {\bibfnamefont {B.}~\bibnamefont
			{Pankovich}}, \bibinfo {author} {\bibfnamefont {A.}~\bibnamefont {Neville}},
		\bibinfo {author} {\bibfnamefont {A.}~\bibnamefont {Kan}}, \bibinfo {author}
		{\bibfnamefont {S.}~\bibnamefont {Omkar}}, \bibinfo {author} {\bibfnamefont
			{K.~H.}\ \bibnamefont {Wan}},\ and\ \bibinfo {author} {\bibfnamefont
			{K.}~\bibnamefont {Br\'adler}},\ }\bibfield  {title} {\bibinfo {title}
		{Flexible entangled-state generation in linear optics},\ }\href
	{https://doi.org/10.1103/PhysRevA.110.032402} {\bibfield  {journal} {\bibinfo
			{journal} {Phys. Rev. A}\ }\textbf {\bibinfo {volume} {110}},\ \bibinfo
		{pages} {032402} (\bibinfo {year} {2024}{\natexlab{a}})}\BibitemShut
	{NoStop}%
	\bibitem [{\citenamefont {Staudacher}\ \emph {et~al.}(2026)\citenamefont
		{Staudacher}, \citenamefont {Jeevanesan},\ and\ \citenamefont
		{Guggemos}}]{Staudacher2026}%
	\BibitemOpen
	\bibfield  {author} {\bibinfo {author} {\bibfnamefont {K.}~\bibnamefont
			{Staudacher}}, \bibinfo {author} {\bibfnamefont {B.}~\bibnamefont
			{Jeevanesan}},\ and\ \bibinfo {author} {\bibfnamefont {T.}~\bibnamefont
			{Guggemos}},\ }\bibfield  {title} {\bibinfo {title} {Adaptive framework for
			failure-aware protocols in fusion-based graph-state generation},\ }\href
	{http://dx.doi.org/10.1103/2z98-6bsp} {\bibfield  {journal} {\bibinfo
			{journal} {Physical Review A}\ }\textbf {\bibinfo {volume} {114}} (\bibinfo
		{year} {2026})}\BibitemShut {NoStop}%
	\bibitem [{\citenamefont {Melkozerov}\ \emph {et~al.}(2026)\citenamefont
		{Melkozerov}, \citenamefont {Straupe},\ and\ \citenamefont
		{Saygin}}]{Melkozerov2026fusion}%
	\BibitemOpen
	\bibfield  {author} {\bibinfo {author} {\bibfnamefont {A.~A.}\ \bibnamefont
			{Melkozerov}}, \bibinfo {author} {\bibfnamefont {S.~S.}\ \bibnamefont
			{Straupe}},\ and\ \bibinfo {author} {\bibfnamefont {M.~{\relax Yu}.}\
			\bibnamefont {Saygin}},\ }\bibfield  {title} {\bibinfo {title}
		{Single-photon-boosted type-{I} fusion gates},\ }\href
	{http://dx.doi.org/10.1103/jy5k-6ptg} {\bibfield  {journal} {\bibinfo
			{journal} {Physical Review Applied}\ }\textbf {\bibinfo {volume} {26}}
		(\bibinfo {year} {2026})}\BibitemShut {NoStop}%
	\bibitem [{\citenamefont {Melkozerov}\ \emph {et~al.}(2024)\citenamefont
		{Melkozerov}, \citenamefont {Avanesov}, \citenamefont {Dyakonov},\ and\
		\citenamefont {Straupe}}]{Melkozerov2024}%
	\BibitemOpen
	\bibfield  {author} {\bibinfo {author} {\bibfnamefont {A.}~\bibnamefont
			{Melkozerov}}, \bibinfo {author} {\bibfnamefont {A.}~\bibnamefont
			{Avanesov}}, \bibinfo {author} {\bibfnamefont {I.}~\bibnamefont {Dyakonov}},\
		and\ \bibinfo {author} {\bibfnamefont {S.}~\bibnamefont {Straupe}},\
	}\bibfield  {title} {\bibinfo {title} {Analysis of optical loss thresholds in
			the fusion-based quantum computing architecture},\ }\href
	{https://doi.org/10.1063/5.0214728} {\bibfield  {journal} {\bibinfo
			{journal} {APL Quantum}\ }\textbf {\bibinfo {volume} {1}},\ \bibinfo {pages}
		{3} (\bibinfo {year} {2024})}\BibitemShut {NoStop}%
	\bibitem [{\citenamefont {Chan}\ \emph {et~al.}(2026)\citenamefont {Chan},
		\citenamefont {Capatos}, \citenamefont {Lodahl}, \citenamefont {Sørensen},\
		and\ \citenamefont {Paesani}}]{Chan2026}%
	\BibitemOpen
	\bibfield  {author} {\bibinfo {author} {\bibfnamefont {M.~L.}\ \bibnamefont
			{Chan}}, \bibinfo {author} {\bibfnamefont {A.~A.}\ \bibnamefont {Capatos}},
		\bibinfo {author} {\bibfnamefont {P.}~\bibnamefont {Lodahl}}, \bibinfo
		{author} {\bibfnamefont {A.~S.}\ \bibnamefont {Sørensen}},\ and\ \bibinfo
		{author} {\bibfnamefont {S.}~\bibnamefont {Paesani}},\ }\bibfield  {title}
	{\bibinfo {title} {Practical blueprint for low-depth photonic quantum
			computing with quantum dots},\ }\href
	{http://dx.doi.org/10.1038/s41534-026-01258-3} {\bibfield  {journal}
		{\bibinfo  {journal} {npj Quantum Information}\ }\textbf {\bibinfo {volume}
			{12}} (\bibinfo {year} {2026})}\BibitemShut {NoStop}%
	\bibitem [{\citenamefont {Calsamiglia}\ and\ \citenamefont
		{L\"{u}tkenhaus}(2001)}]{Calsamiglia2001}%
	\BibitemOpen
	\bibfield  {author} {\bibinfo {author} {\bibfnamefont {J.}~\bibnamefont
			{Calsamiglia}}\ and\ \bibinfo {author} {\bibfnamefont {N.}~\bibnamefont
			{L\"{u}tkenhaus}},\ }\bibfield  {title} {\bibinfo {title} {Maximum efficiency
			of a linear-optical bell-state analyzer},\ }\href
	{https://doi.org/10.1007/s003400000484} {\bibfield  {journal} {\bibinfo
			{journal} {Applied Physics B}\ }\textbf {\bibinfo {volume} {72}},\ \bibinfo
		{pages} {67–71} (\bibinfo {year} {2001})}\BibitemShut {NoStop}%
	\bibitem [{\citenamefont {Grice}(2011)}]{Grice2011}%
	\BibitemOpen
	\bibfield  {author} {\bibinfo {author} {\bibfnamefont {W.~P.}\ \bibnamefont
			{Grice}},\ }\bibfield  {title} {\bibinfo {title} {Arbitrarily complete
			bell-state measurement using only linear optical elements},\ }\href
	{https://doi.org/10.1103/physreva.84.042331} {\bibfield  {journal} {\bibinfo
			{journal} {Physical Review A}\ }\textbf {\bibinfo {volume} {84}},\ \bibinfo
		{pages} {4} (\bibinfo {year} {2011})}\BibitemShut {NoStop}%
	\bibitem [{\citenamefont {Ewert}\ and\ \citenamefont {van
			Loock}(2014)}]{Ewert2014}%
	\BibitemOpen
	\bibfield  {author} {\bibinfo {author} {\bibfnamefont {F.}~\bibnamefont
			{Ewert}}\ and\ \bibinfo {author} {\bibfnamefont {P.}~\bibnamefont {van
				Loock}},\ }\bibfield  {title} {\bibinfo {title} {$3/4$-efficient bell
			measurement with passive linear optics and unentangled ancillae},\ }\href
	{https://doi.org/10.1103/physrevlett.113.140403} {\bibfield  {journal}
		{\bibinfo  {journal} {Physical Review Letters}\ }\textbf {\bibinfo {volume}
			{113}},\ \bibinfo {pages} {14} (\bibinfo {year} {2014})}\BibitemShut
	{NoStop}%
	\bibitem [{\citenamefont {Hauser}\ \emph {et~al.}(2025)\citenamefont {Hauser},
		\citenamefont {Bayerbach}, \citenamefont {D’Aurelio}, \citenamefont
		{Weber}, \citenamefont {Santandrea} \emph {et~al.}}]{Hauser2025}%
	\BibitemOpen
	\bibfield  {author} {\bibinfo {author} {\bibfnamefont {N.}~\bibnamefont
			{Hauser}}, \bibinfo {author} {\bibfnamefont {M.~J.}\ \bibnamefont
			{Bayerbach}}, \bibinfo {author} {\bibfnamefont {S.~E.}\ \bibnamefont
			{D’Aurelio}}, \bibinfo {author} {\bibfnamefont {R.}~\bibnamefont {Weber}},
		\bibinfo {author} {\bibfnamefont {M.}~\bibnamefont {Santandrea}}, \emph
		{et~al.},\ }\bibfield  {title} {\bibinfo {title} {Boosted bell-state
			measurements for photonic quantum computation},\ }\href
	{https://doi.org/10.1038/s41534-025-00986-2} {\bibfield  {journal} {\bibinfo
			{journal} {npj Quantum Information}\ }\textbf {\bibinfo {volume} {11}},\
		\bibinfo {pages} {1} (\bibinfo {year} {2025})}\BibitemShut {NoStop}%
	\bibitem [{\citenamefont {Bayerbach}\ \emph {et~al.}(2023)\citenamefont
		{Bayerbach}, \citenamefont {D’Aurelio}, \citenamefont {van Loock},\ and\
		\citenamefont {Barz}}]{Bayerbach2023}%
	\BibitemOpen
	\bibfield  {author} {\bibinfo {author} {\bibfnamefont {M.~J.}\ \bibnamefont
			{Bayerbach}}, \bibinfo {author} {\bibfnamefont {S.~E.}\ \bibnamefont
			{D’Aurelio}}, \bibinfo {author} {\bibfnamefont {P.}~\bibnamefont {van
				Loock}},\ and\ \bibinfo {author} {\bibfnamefont {S.}~\bibnamefont {Barz}},\
	}\bibfield  {title} {\bibinfo {title} {Bell-state measurement exceeding 50\%
			success probability with linear optics},\ }\href
	{https://doi.org/10.1126/sciadv.adf4080} {\bibfield  {journal} {\bibinfo
			{journal} {Science Advances}\ }\textbf {\bibinfo {volume} {9}},\ \bibinfo
		{pages} {32} (\bibinfo {year} {2023})}\BibitemShut {NoStop}%
	\bibitem [{\citenamefont {Guo}\ \emph {et~al.}(2026)\citenamefont {Guo},
		\citenamefont {Zou}, \citenamefont {Ding}, \citenamefont {Zhang},
		\citenamefont {Xu}, \citenamefont {Liu}, \citenamefont {Zhao}, \citenamefont
		{Ge}, \citenamefont {Peng}, \citenamefont {Xu} \emph {et~al.}}]{Guo2026}%
	\BibitemOpen
	\bibfield  {author} {\bibinfo {author} {\bibfnamefont {Y.-P.}\ \bibnamefont
			{Guo}}, \bibinfo {author} {\bibfnamefont {G.-Y.}\ \bibnamefont {Zou}},
		\bibinfo {author} {\bibfnamefont {X.}~\bibnamefont {Ding}}, \bibinfo {author}
		{\bibfnamefont {Q.-H.}\ \bibnamefont {Zhang}}, \bibinfo {author}
		{\bibfnamefont {M.-C.}\ \bibnamefont {Xu}}, \bibinfo {author} {\bibfnamefont
			{R.-Z.}\ \bibnamefont {Liu}}, \bibinfo {author} {\bibfnamefont {J.-Y.}\
			\bibnamefont {Zhao}}, \bibinfo {author} {\bibfnamefont {Z.-X.}\ \bibnamefont
			{Ge}}, \bibinfo {author} {\bibfnamefont {L.-C.}\ \bibnamefont {Peng}},
		\bibinfo {author} {\bibfnamefont {K.-M.}\ \bibnamefont {Xu}}, \emph
		{et~al.},\ }\bibfield  {title} {\bibinfo {title} {Boosted fusion gates above
			the percolation threshold for scalable graph-state generation},\ }\href
	{http://dx.doi.org/10.1103/xt7c-qv52} {\bibfield  {journal} {\bibinfo
			{journal} {Physical Review A}\ }\textbf {\bibinfo {volume} {113}} (\bibinfo
		{year} {2026})}\BibitemShut {NoStop}%
	\bibitem [{\citenamefont {Stanisic}\ \emph {et~al.}(2017)\citenamefont
		{Stanisic}, \citenamefont {Linden}, \citenamefont {Montanaro},\ and\
		\citenamefont {Turner}}]{Stanisic2017}%
	\BibitemOpen
	\bibfield  {author} {\bibinfo {author} {\bibfnamefont {S.}~\bibnamefont
			{Stanisic}}, \bibinfo {author} {\bibfnamefont {N.}~\bibnamefont {Linden}},
		\bibinfo {author} {\bibfnamefont {A.}~\bibnamefont {Montanaro}},\ and\
		\bibinfo {author} {\bibfnamefont {P.~S.}\ \bibnamefont {Turner}},\ }\bibfield
	{title} {\bibinfo {title} {Generating entanglement with linear optics},\
	}\href {https://doi.org/10.1103/physreva.96.043861} {\bibfield  {journal}
		{\bibinfo  {journal} {Physical Review A}\ }\textbf {\bibinfo {volume} {96}},\
		\bibinfo {pages} {4} (\bibinfo {year} {2017})}\BibitemShut {NoStop}%
	\bibitem [{\citenamefont {Wein}\ \emph {et~al.}(2025)\citenamefont {Wein},
		\citenamefont {Goubault~de Brugière}, \citenamefont {Music}, \citenamefont
		{Senellart}, \citenamefont {Bourdoncle},\ and\ \citenamefont
		{Mansfield}}]{Wein2025}%
	\BibitemOpen
	\bibfield  {author} {\bibinfo {author} {\bibfnamefont {S.~C.}\ \bibnamefont
			{Wein}}, \bibinfo {author} {\bibfnamefont {T.}~\bibnamefont {Goubault~de
				Brugière}}, \bibinfo {author} {\bibfnamefont {L.}~\bibnamefont {Music}},
		\bibinfo {author} {\bibfnamefont {P.}~\bibnamefont {Senellart}}, \bibinfo
		{author} {\bibfnamefont {B.}~\bibnamefont {Bourdoncle}},\ and\ \bibinfo
		{author} {\bibfnamefont {S.}~\bibnamefont {Mansfield}},\ }\bibfield  {title}
	{\bibinfo {title} {Minimizing resource overhead in fusion-based quantum
			computation using hybrid spin-photon devices},\ }\href
	{https://doi.org/10.1103/8l5g-x3b7} {\bibfield  {journal} {\bibinfo
			{journal} {PRX Quantum}\ }\textbf {\bibinfo {volume} {6}},\ \bibinfo {pages}
		{4} (\bibinfo {year} {2025})}\BibitemShut {NoStop}%
	\bibitem [{\citenamefont {Bartolucci}\ \emph
		{et~al.}(2021{\natexlab{b}})\citenamefont {Bartolucci}, \citenamefont
		{Birchall}, \citenamefont {Bonneau}, \citenamefont {Cable}, \citenamefont
		{Gimeno-Segovia}, \citenamefont {Kieling}, \citenamefont {Nickerson},
		\citenamefont {Rudolph},\ and\ \citenamefont
		{Sparrow}}]{Switch_networks2021}%
	\BibitemOpen
	\bibfield  {author} {\bibinfo {author} {\bibfnamefont {S.}~\bibnamefont
			{Bartolucci}}, \bibinfo {author} {\bibfnamefont {P.}~\bibnamefont
			{Birchall}}, \bibinfo {author} {\bibfnamefont {D.}~\bibnamefont {Bonneau}},
		\bibinfo {author} {\bibfnamefont {H.}~\bibnamefont {Cable}}, \bibinfo
		{author} {\bibfnamefont {M.}~\bibnamefont {Gimeno-Segovia}}, \bibinfo
		{author} {\bibfnamefont {K.}~\bibnamefont {Kieling}}, \bibinfo {author}
		{\bibfnamefont {N.}~\bibnamefont {Nickerson}}, \bibinfo {author}
		{\bibfnamefont {T.}~\bibnamefont {Rudolph}},\ and\ \bibinfo {author}
		{\bibfnamefont {C.}~\bibnamefont {Sparrow}},\ }\href@noop {} {\bibinfo
		{title} {Switch networks for photonic fusion-based quantum computing}}
	(\bibinfo {year} {2021}{\natexlab{b}}),\ \Eprint
	{https://arxiv.org/abs/2109.13760} {arXiv:2109.13760} \BibitemShut {NoStop}%
	\bibitem [{\citenamefont {Forbes}\ \emph {et~al.}(2025)\citenamefont {Forbes},
		\citenamefont {Ghafari}, \citenamefont {Deacon}, \citenamefont {Singh},
		\citenamefont {Lavie} \emph {et~al.}}]{Forbes2025}%
	\BibitemOpen
	\bibfield  {author} {\bibinfo {author} {\bibfnamefont {I.}~\bibnamefont
			{Forbes}}, \bibinfo {author} {\bibfnamefont {F.}~\bibnamefont {Ghafari}},
		\bibinfo {author} {\bibfnamefont {E.~C.~R.}\ \bibnamefont {Deacon}}, \bibinfo
		{author} {\bibfnamefont {S.~P.}\ \bibnamefont {Singh}}, \bibinfo {author}
		{\bibfnamefont {E.}~\bibnamefont {Lavie}}, \emph {et~al.},\ }\bibfield
	{title} {\bibinfo {title} {Heralded generation of entanglement with
			photons},\ }\href {https://doi.org/10.1088/1361-6633/adf85e} {\bibfield
		{journal} {\bibinfo  {journal} {Reports on Progress in Physics}\ }\textbf
		{\bibinfo {volume} {88}},\ \bibinfo {pages} {086002} (\bibinfo {year}
		{2025})}\BibitemShut {NoStop}%
	\bibitem [{\citenamefont {Barz}\ \emph {et~al.}(2010)\citenamefont {Barz},
		\citenamefont {Cronenberg}, \citenamefont {Zeilinger},\ and\ \citenamefont
		{Walther}}]{Barz2010}%
	\BibitemOpen
	\bibfield  {author} {\bibinfo {author} {\bibfnamefont {S.}~\bibnamefont
			{Barz}}, \bibinfo {author} {\bibfnamefont {G.}~\bibnamefont {Cronenberg}},
		\bibinfo {author} {\bibfnamefont {A.}~\bibnamefont {Zeilinger}},\ and\
		\bibinfo {author} {\bibfnamefont {P.}~\bibnamefont {Walther}},\ }\bibfield
	{title} {\bibinfo {title} {Heralded generation of entangled photon pairs},\
	}\href {https://doi.org/10.1038/nphoton.2010.156} {\bibfield  {journal}
		{\bibinfo  {journal} {Nature Photonics}\ }\textbf {\bibinfo {volume} {4}},\
		\bibinfo {pages} {553–556} (\bibinfo {year} {2010})}\BibitemShut {NoStop}%
	\bibitem [{\citenamefont {Gubarev}\ \emph {et~al.}(2020)\citenamefont
		{Gubarev}, \citenamefont {Dyakonov}, \citenamefont {Saygin}, \citenamefont
		{Struchalin}, \citenamefont {Straupe},\ and\ \citenamefont
		{Kulik}}]{Gubarev2020}%
	\BibitemOpen
	\bibfield  {author} {\bibinfo {author} {\bibfnamefont {F.~V.}\ \bibnamefont
			{Gubarev}}, \bibinfo {author} {\bibfnamefont {I.~V.}\ \bibnamefont
			{Dyakonov}}, \bibinfo {author} {\bibfnamefont {M.~{\relax Yu}.}\ \bibnamefont
			{Saygin}}, \bibinfo {author} {\bibfnamefont {G.~I.}\ \bibnamefont
			{Struchalin}}, \bibinfo {author} {\bibfnamefont {S.~S.}\ \bibnamefont
			{Straupe}},\ and\ \bibinfo {author} {\bibfnamefont {S.~P.}\ \bibnamefont
			{Kulik}},\ }\bibfield  {title} {\bibinfo {title} {Improved heralded schemes
			to generate entangled states from single photons},\ }\href
	{https://doi.org/10.1103/physreva.102.012604} {\bibfield  {journal} {\bibinfo
			{journal} {Physical Review A}\ }\textbf {\bibinfo {volume} {102}},\ \bibinfo
		{pages} {1} (\bibinfo {year} {2020})}\BibitemShut {NoStop}%
	\bibitem [{\citenamefont {Fldzhyan}\ \emph {et~al.}(2021)\citenamefont
		{Fldzhyan}, \citenamefont {Saygin},\ and\ \citenamefont
		{Kulik}}]{Fldzhyan2021}%
	\BibitemOpen
	\bibfield  {author} {\bibinfo {author} {\bibfnamefont {S.~A.}\ \bibnamefont
			{Fldzhyan}}, \bibinfo {author} {\bibfnamefont {M.~{\relax Yu}.}\ \bibnamefont
			{Saygin}},\ and\ \bibinfo {author} {\bibfnamefont {S.~P.}\ \bibnamefont
			{Kulik}},\ }\bibfield  {title} {\bibinfo {title} {Compact linear optical
			scheme for bell state generation},\ }\href
	{https://doi.org/10.1103/physrevresearch.3.043031} {\bibfield  {journal}
		{\bibinfo  {journal} {Physical Review Research}\ }\textbf {\bibinfo {volume}
			{3}},\ \bibinfo {pages} {4} (\bibinfo {year} {2021})}\BibitemShut {NoStop}%
	\bibitem [{\citenamefont {Skryabin}\ \emph {et~al.}(2025)\citenamefont
		{Skryabin}, \citenamefont {Biriukov}, \citenamefont {Dryazgov}, \citenamefont
		{Fldzhyan}, \citenamefont {Zhuravitskii} \emph {et~al.}}]{Skryabin2025}%
	\BibitemOpen
	\bibfield  {author} {\bibinfo {author} {\bibfnamefont {N.~N.}\ \bibnamefont
			{Skryabin}}, \bibinfo {author} {\bibfnamefont {Y.~A.}\ \bibnamefont
			{Biriukov}}, \bibinfo {author} {\bibfnamefont {M.~A.}\ \bibnamefont
			{Dryazgov}}, \bibinfo {author} {\bibfnamefont {S.~A.}\ \bibnamefont
			{Fldzhyan}}, \bibinfo {author} {\bibfnamefont {S.~A.}\ \bibnamefont
			{Zhuravitskii}}, \emph {et~al.},\ }\bibfield  {title} {\bibinfo {title}
		{Heralded generation of programmable two-qubit entangled states on a
			linear-optical platform},\ }\href {https://doi.org/10.1364/opticaq.546244}
	{\bibfield  {journal} {\bibinfo  {journal} {Optica Quantum}\ }\textbf
		{\bibinfo {volume} {3}},\ \bibinfo {pages} {162} (\bibinfo {year}
		{2025})}\BibitemShut {NoStop}%
	\bibitem [{\citenamefont {Bell}\ \emph {et~al.}(2014)\citenamefont {Bell},
		\citenamefont {Markham}, \citenamefont {Herrera-Martí}, \citenamefont
		{Marin}, \citenamefont {Wadsworth}, \citenamefont {Rarity},\ and\
		\citenamefont {Tame}}]{Bell2014}%
	\BibitemOpen
	\bibfield  {author} {\bibinfo {author} {\bibfnamefont {B.~A.}\ \bibnamefont
			{Bell}}, \bibinfo {author} {\bibfnamefont {D.}~\bibnamefont {Markham}},
		\bibinfo {author} {\bibfnamefont {D.~A.}\ \bibnamefont {Herrera-Martí}},
		\bibinfo {author} {\bibfnamefont {A.}~\bibnamefont {Marin}}, \bibinfo
		{author} {\bibfnamefont {W.~J.}\ \bibnamefont {Wadsworth}}, \bibinfo {author}
		{\bibfnamefont {J.~G.}\ \bibnamefont {Rarity}},\ and\ \bibinfo {author}
		{\bibfnamefont {M.~S.}\ \bibnamefont {Tame}},\ }\bibfield  {title} {\bibinfo
		{title} {Experimental demonstration of graph-state quantum secret sharing},\
	}\href {http://dx.doi.org/10.1038/ncomms6480} {\bibfield  {journal} {\bibinfo
			{journal} {Nature Communications}\ }\textbf {\bibinfo {volume} {5}}
		(\bibinfo {year} {2014})}\BibitemShut {NoStop}%
	\bibitem [{\citenamefont {de~Jong}\ \emph {et~al.}(2023)\citenamefont
		{de~Jong}, \citenamefont {Hahn}, \citenamefont {Eisert}, \citenamefont
		{Walk},\ and\ \citenamefont {Pappa}}]{deJong2023}%
	\BibitemOpen
	\bibfield  {author} {\bibinfo {author} {\bibfnamefont {J.}~\bibnamefont
			{de~Jong}}, \bibinfo {author} {\bibfnamefont {F.}~\bibnamefont {Hahn}},
		\bibinfo {author} {\bibfnamefont {J.}~\bibnamefont {Eisert}}, \bibinfo
		{author} {\bibfnamefont {N.}~\bibnamefont {Walk}},\ and\ \bibinfo {author}
		{\bibfnamefont {A.}~\bibnamefont {Pappa}},\ }\bibfield  {title} {\bibinfo
		{title} {Anonymous conference key agreement in linear quantum networks},\
	}\href {https://doi.org/10.22331/q-2023-09-21-1117} {\bibfield  {journal}
		{\bibinfo  {journal} {Quantum}\ }\textbf {\bibinfo {volume} {7}},\ \bibinfo
		{pages} {1117} (\bibinfo {year} {2023})}\BibitemShut {NoStop}%
	\bibitem [{\citenamefont {Raussendorf}(2003)}]{Raussendorf2003}%
	\BibitemOpen
	\bibfield  {author} {\bibinfo {author} {\bibfnamefont {R.}~\bibnamefont
			{Raussendorf}},\ }\emph {\bibinfo {title} {Measurement-based quantum
			computation with cluster states}},\ \href@noop {} {Ph.D. thesis},\ \bibinfo
	{school} {Ludwig-Maximilians-Universit\"{a}t M\"{u}nchen} (\bibinfo {year}
	{2003})\BibitemShut {NoStop}%
	\bibitem [{\citenamefont {Sahay}\ \emph {et~al.}(2023)\citenamefont {Sahay},
		\citenamefont {Claes},\ and\ \citenamefont {Puri}}]{Sahay2023}%
	\BibitemOpen
	\bibfield  {author} {\bibinfo {author} {\bibfnamefont {K.}~\bibnamefont
			{Sahay}}, \bibinfo {author} {\bibfnamefont {J.}~\bibnamefont {Claes}},\ and\
		\bibinfo {author} {\bibfnamefont {S.}~\bibnamefont {Puri}},\ }\bibfield
	{title} {\bibinfo {title} {Tailoring fusion-based error correction for high
			thresholds to biased fusion failures},\ }\href
	{https://doi.org/10.1103/physrevlett.131.120604} {\bibfield  {journal}
		{\bibinfo  {journal} {Physical Review Letters}\ }\textbf {\bibinfo {volume}
			{131}},\ \bibinfo {pages} {12} (\bibinfo {year} {2023})}\BibitemShut
	{NoStop}%
	\bibitem [{\citenamefont {Varnava}\ \emph {et~al.}(2008)\citenamefont
		{Varnava}, \citenamefont {Browne},\ and\ \citenamefont
		{Rudolph}}]{Varnava2008}%
	\BibitemOpen
	\bibfield  {author} {\bibinfo {author} {\bibfnamefont {M.}~\bibnamefont
			{Varnava}}, \bibinfo {author} {\bibfnamefont {D.~E.}\ \bibnamefont
			{Browne}},\ and\ \bibinfo {author} {\bibfnamefont {T.}~\bibnamefont
			{Rudolph}},\ }\bibfield  {title} {\bibinfo {title} {How good must single
			photon sources and detectors be for efficient linear optical quantum
			computation?},\ }\href {https://doi.org/10.1103/physrevlett.100.060502}
	{\bibfield  {journal} {\bibinfo  {journal} {Physical Review Letters}\
		}\textbf {\bibinfo {volume} {100}},\ \bibinfo {pages} {6} (\bibinfo {year}
		{2008})}\BibitemShut {NoStop}%
	\bibitem [{\citenamefont {Bhatti}\ and\ \citenamefont
		{Barz}(2025)}]{Bhatti2025}%
	\BibitemOpen
	\bibfield  {author} {\bibinfo {author} {\bibfnamefont {D.}~\bibnamefont
			{Bhatti}}\ and\ \bibinfo {author} {\bibfnamefont {S.}~\bibnamefont {Barz}},\
	}\bibfield  {title} {\bibinfo {title} {Heralding higher-dimensional bell and
			greenberger–horne–zeilinger states using multiport splitters},\ }\href
	{https://doi.org/10.1088/1367-2630/adb903} {\bibfield  {journal} {\bibinfo
			{journal} {New Journal of Physics}\ }\textbf {\bibinfo {volume} {27}},\
		\bibinfo {pages} {033006} (\bibinfo {year} {2025})}\BibitemShut {NoStop}%
	\bibitem [{\citenamefont {Gimeno-Segovia}(2016)}]{gimeno2016}%
	\BibitemOpen
	\bibfield  {author} {\bibinfo {author} {\bibfnamefont {M.}~\bibnamefont
			{Gimeno-Segovia}},\ }\emph {\bibinfo {title} {Towards practical linear
			optical quantum computing}},\ \href@noop {} {Ph.D. thesis},\ \bibinfo
	{school} {Imperial College London} (\bibinfo {year} {2016})\BibitemShut
	{NoStop}%
	\bibitem [{\citenamefont {Pankovich}\ \emph
		{et~al.}(2024{\natexlab{b}})\citenamefont {Pankovich}, \citenamefont {Kan},
		\citenamefont {Wan}, \citenamefont {Ostmann}, \citenamefont {Neville} \emph
		{et~al.}}]{Pankovich2024}%
	\BibitemOpen
	\bibfield  {author} {\bibinfo {author} {\bibfnamefont {B.}~\bibnamefont
			{Pankovich}}, \bibinfo {author} {\bibfnamefont {A.}~\bibnamefont {Kan}},
		\bibinfo {author} {\bibfnamefont {K.~H.}\ \bibnamefont {Wan}}, \bibinfo
		{author} {\bibfnamefont {M.}~\bibnamefont {Ostmann}}, \bibinfo {author}
		{\bibfnamefont {A.}~\bibnamefont {Neville}}, \emph {et~al.},\ }\bibfield
	{title} {\bibinfo {title} {High-photon-loss threshold quantum computing using
			ghz-state measurements},\ }\href
	{https://doi.org/10.1103/physrevlett.133.050604} {\bibfield  {journal}
		{\bibinfo  {journal} {Physical Review Letters}\ }\textbf {\bibinfo {volume}
			{133}},\ \bibinfo {pages} {5} (\bibinfo {year}
		{2024}{\natexlab{b}})}\BibitemShut {NoStop}%
	\bibitem [{\citenamefont {Patil}\ and\ \citenamefont {Guha}(2026)}]{Patil2026}%
	\BibitemOpen
	\bibfield  {author} {\bibinfo {author} {\bibfnamefont {A.}~\bibnamefont
			{Patil}}\ and\ \bibinfo {author} {\bibfnamefont {S.}~\bibnamefont {Guha}},\
	}\bibfield  {title} {\bibinfo {title} {A graphical rule book for clifford
			manipulations of stabilizer states},\ }\href
	{https://doi.org/10.1109/tqe.2026.3653200} {\bibfield  {journal} {\bibinfo
			{journal} {IEEE Transactions on Quantum Engineering}\ }\textbf {\bibinfo
			{volume} {7}},\ \bibinfo {pages} {1–33} (\bibinfo {year}
		{2026})}\BibitemShut {NoStop}%
	\bibitem [{\citenamefont {Lee}\ \emph {et~al.}(2015)\citenamefont {Lee},
		\citenamefont {Park}, \citenamefont {Ralph},\ and\ \citenamefont
		{Jeong}}]{Lee2015}%
	\BibitemOpen
	\bibfield  {author} {\bibinfo {author} {\bibfnamefont {S.-W.}\ \bibnamefont
			{Lee}}, \bibinfo {author} {\bibfnamefont {K.}~\bibnamefont {Park}}, \bibinfo
		{author} {\bibfnamefont {T.~C.}\ \bibnamefont {Ralph}},\ and\ \bibinfo
		{author} {\bibfnamefont {H.}~\bibnamefont {Jeong}},\ }\bibfield  {title}
	{\bibinfo {title} {Nearly deterministic bell measurement for multiphoton
			qubits and its application to quantum information processing},\ }\href
	{http://dx.doi.org/10.1103/PhysRevLett.114.113603} {\bibfield  {journal}
		{\bibinfo  {journal} {Physical Review Letters}\ }\textbf {\bibinfo {volume}
			{114}} (\bibinfo {year} {2015})}\BibitemShut {NoStop}%
	\bibitem [{\citenamefont {Ewert}\ \emph {et~al.}(2016)\citenamefont {Ewert},
		\citenamefont {Bergmann},\ and\ \citenamefont {van Loock}}]{Ewert2016}%
	\BibitemOpen
	\bibfield  {author} {\bibinfo {author} {\bibfnamefont {F.}~\bibnamefont
			{Ewert}}, \bibinfo {author} {\bibfnamefont {M.}~\bibnamefont {Bergmann}},\
		and\ \bibinfo {author} {\bibfnamefont {P.}~\bibnamefont {van Loock}},\
	}\bibfield  {title} {\bibinfo {title} {Ultrafast long-distance quantum
			communication with static linear optics},\ }\href
	{http://dx.doi.org/10.1103/PhysRevLett.117.210501} {\bibfield  {journal}
		{\bibinfo  {journal} {Physical Review Letters}\ }\textbf {\bibinfo {volume}
			{117}} (\bibinfo {year} {2016})}\BibitemShut {NoStop}%
	\bibitem [{\citenamefont {Kumar}\ \emph {et~al.}(2026)\citenamefont {Kumar},
		\citenamefont {Reiß}, \citenamefont {van Loock},\ and\ \citenamefont
		{Barz}}]{Kumar2026}%
	\BibitemOpen
	\bibfield  {author} {\bibinfo {author} {\bibfnamefont {S.}~\bibnamefont
			{Kumar}}, \bibinfo {author} {\bibfnamefont {S.~D.}\ \bibnamefont {Reiß}},
		\bibinfo {author} {\bibfnamefont {P.}~\bibnamefont {van Loock}},\ and\
		\bibinfo {author} {\bibfnamefont {S.}~\bibnamefont {Barz}},\ }\href@noop {}
	{\bibinfo {title} {Demonstration of a logical bell-state measurement beyond
			the linear-optical limit}} (\bibinfo {year} {2026}),\ \Eprint
	{https://arxiv.org/abs/2609.10698} {arXiv:2609.10698} \BibitemShut {NoStop}%
	\bibitem [{\citenamefont {Collins}\ \emph {et~al.}(2013)\citenamefont
		{Collins}, \citenamefont {Xiong}, \citenamefont {Rey}, \citenamefont {Vo},
		\citenamefont {He}, \citenamefont {Shahnia}, \citenamefont {Reardon},
		\citenamefont {Krauss}, \citenamefont {Steel}, \citenamefont {Clark},\ and\
		\citenamefont {Eggleton}}]{Collins2013}%
	\BibitemOpen
	\bibfield  {author} {\bibinfo {author} {\bibfnamefont {M.}~\bibnamefont
			{Collins}}, \bibinfo {author} {\bibfnamefont {C.}~\bibnamefont {Xiong}},
		\bibinfo {author} {\bibfnamefont {I.}~\bibnamefont {Rey}}, \bibinfo {author}
		{\bibfnamefont {T.}~\bibnamefont {Vo}}, \bibinfo {author} {\bibfnamefont
			{J.}~\bibnamefont {He}}, \bibinfo {author} {\bibfnamefont {S.}~\bibnamefont
			{Shahnia}}, \bibinfo {author} {\bibfnamefont {C.}~\bibnamefont {Reardon}},
		\bibinfo {author} {\bibfnamefont {T.}~\bibnamefont {Krauss}}, \bibinfo
		{author} {\bibfnamefont {M.}~\bibnamefont {Steel}}, \bibinfo {author}
		{\bibfnamefont {A.}~\bibnamefont {Clark}},\ and\ \bibinfo {author}
		{\bibfnamefont {B.}~\bibnamefont {Eggleton}},\ }\bibfield  {title} {\bibinfo
		{title} {Integrated spatial multiplexing of heralded single-photon sources},\
	}\href {http://dx.doi.org/10.1038/ncomms3582} {\bibfield  {journal} {\bibinfo
			{journal} {Nature Communications}\ }\textbf {\bibinfo {volume} {4}}
		(\bibinfo {year} {2013})}\BibitemShut {NoStop}%
	\bibitem [{\citenamefont {Kaneda}\ and\ \citenamefont
		{Yabuno}(2026)}]{Kaneda2026}%
	\BibitemOpen
	\bibfield  {author} {\bibinfo {author} {\bibfnamefont {F.}~\bibnamefont
			{Kaneda}}\ and\ \bibinfo {author} {\bibfnamefont {M.}~\bibnamefont
			{Yabuno}},\ }\bibfield  {title} {\bibinfo {title} {Active multiplexing for
			scalable generation and manipulation of photonic quantum states},\ }\href
	{https://doi.org/10.1186/s40580-026-00540-6} {\bibfield  {journal} {\bibinfo
			{journal} {Nano Convergence}\ }\textbf {\bibinfo {volume} {13}},\ \bibinfo
		{pages} {1} (\bibinfo {year} {2026})}\BibitemShut {NoStop}%
	\bibitem [{\citenamefont {Liu}\ \emph {et~al.}(2024)\citenamefont {Liu},
		\citenamefont {Qiao}, \citenamefont {Lachman}, \citenamefont {Ge},
		\citenamefont {Chung}, \citenamefont {Zhao}, \citenamefont {Li},
		\citenamefont {You}, \citenamefont {Filip},\ and\ \citenamefont
		{Huo}}]{Liu2024}%
	\BibitemOpen
	\bibfield  {author} {\bibinfo {author} {\bibfnamefont {R.-Z.}\ \bibnamefont
			{Liu}}, \bibinfo {author} {\bibfnamefont {Y.-K.}\ \bibnamefont {Qiao}},
		\bibinfo {author} {\bibfnamefont {L.}~\bibnamefont {Lachman}}, \bibinfo
		{author} {\bibfnamefont {Z.-X.}\ \bibnamefont {Ge}}, \bibinfo {author}
		{\bibfnamefont {T.-H.}\ \bibnamefont {Chung}}, \bibinfo {author}
		{\bibfnamefont {J.-Y.}\ \bibnamefont {Zhao}}, \bibinfo {author}
		{\bibfnamefont {H.}~\bibnamefont {Li}}, \bibinfo {author} {\bibfnamefont
			{L.}~\bibnamefont {You}}, \bibinfo {author} {\bibfnamefont {R.}~\bibnamefont
			{Filip}},\ and\ \bibinfo {author} {\bibfnamefont {Y.-H.}\ \bibnamefont
			{Huo}},\ }\bibfield  {title} {\bibinfo {title} {Experimental quantum
			non-gaussian coincidences of entangled photons},\ }\href
	{http://dx.doi.org/10.1103/PhysRevLett.132.083601} {\bibfield  {journal}
		{\bibinfo  {journal} {Physical Review Letters}\ }\textbf {\bibinfo {volume}
			{132}} (\bibinfo {year} {2024})}\BibitemShut {NoStop}%
\end{thebibliography}
\end{document}